%% file: main.tex
\documentclass[10pt,aps,pra,superscriptaddress,floatfix,nofootinbib]{revtex4-2}
\usepackage{graphicx}
\usepackage{times}
\usepackage{geometry}
\usepackage{rotating}
\usepackage{multirow}
\usepackage{makecell}
\usepackage[utf8]{inputenc}
\usepackage{amssymb}
\usepackage{amsmath}
\usepackage{ragged2e}
\usepackage{footmisc}
\usepackage{comment}
\usepackage{upgreek}
\usepackage{multirow}
\usepackage{booktabs}
\usepackage{datetime}
\usepackage[dvipsnames]{xcolor} 
\definecolor{xlinkcolor}{cmyk}{1,1,0,0}
\usepackage{url}
\usepackage[
 colorlinks=true,    
 linkcolor=xlinkcolor,     
 citecolor=xlinkcolor,     
 filecolor=xlinkcolor,  
 urlcolor=xlinkcolor,      
 final=true
]{hyperref}
\usepackage{enumitem}
\usepackage[left]{lineno}

\setenumerate{itemsep=0mm}
\input{defs.tex}

\begin{document}
\title{Review of Neutron Yield from \texorpdfstring{\alphan}{(alpha,n)} Reactions: Data, Methods, and Prospects}


\input{authors.tex}
\begin{abstract}
  
Understanding the radiogenic neutron production rate through the \alphan\ reaction is crucial in many areas of physics, including dark matter searches, neutrino studies, and nuclear astrophysics. In addition to its relevance for fundamental research, the \alphan\ reaction also plays a significant role in nuclear energy technologies, for example by contributing to neutron production in subcritical systems using UO$_{2}$, as well as in applications such as medical physics. This review examines the current state of \alphan\ yield calculations and neutron spectra, describes the computational tools used for their estimation, and discusses the available cross-section data. We explore the uncertainties affecting \alphan\  yield estimations and propose a strategy to enhance their accuracy. Furthermore, this paper discusses and emphasizes the need for new measurements of \alphan\ cross-sections for a variety of relevant materials. Such measurements are essential for improving neutron flux predictions, which are crucial for reducing uncertainties in sensitivity estimates for next-generation physics experiments operating in the keV–-MeV range.

\end{abstract}

\maketitle
\tableofcontents

\input{TexFiles/Introduction}
\input{TexFiles/Process}
\input{TexFiles/Isotopes}
\input{TexFiles/Data}
\input{TexFiles/Tools}
\input{TexFiles/Uncertanties}

\input{TexFiles/Measurements}

\input{TexFiles/Conclusions}

\input{TexFiles/Acknowledgments}

\bibliographystyle{apsrev4-2}
\bibliography{biblio}

\end{document}

%% file: defs.tex
\usepackage[utf8]{inputenc}
\usepackage{charter}
\usepackage{graphicx}
\usepackage{lipsum}
\usepackage{amsmath}
\usepackage[version=4]{mhchem}

\usepackage[
  separate-uncertainty,
  per-mode=symbol,
  range-phrase=--,
  range-units=single
]{siunitx}

\usepackage{enumitem}
\usepackage{todonotes}
\usepackage{amssymb}
\DeclareSIUnit\molar{M}
\DeclareSIUnit\c{\mbox{$c$}}
\DeclareSIUnit\week{w}
\DeclareSIUnit\year{yr}
\DeclareSIUnit\yr{yr}
\DeclareSIUnit\yr{yr}
\DeclareSIUnit\standard{std}
\DeclareSIUnit\str{sr}
\DeclareSIUnit\MV{MV}
\DeclareSIUnit\ppm{ppm}
\DeclareSIUnit\ppb{ppb}
\DeclareSIUnit\ppt{ppt}
\DeclareSIUnit\pe{PE}
\DeclareSIUnit\spe{SPE}
\DeclareSIUnit\ev{events}
\DeclareSIUnit\hit{hit}
\DeclareSIUnit\hits{hits}
\DeclareSIUnit\bin{(\mbox{5-PE}~bin)}
\DeclareSIUnit\sgm{\mbox{$\sigma$}}
\DeclareSIUnit\rms{RMS}
\DeclareSIUnit\keVr{\mbox{keV$_{\rm nr}$}}
\DeclareSIUnit\keVee{\mbox{keV$_{\rm ee}$}}
\DeclareSIUnit\ph{photons}
\DeclareSIUnit\pm{PMT}
\DeclareSIUnit\inch{''}
\DeclareSIUnit\bit{bit}
\DeclareSIUnit\sample{S}
\DeclareSIUnit\barn{b}
\DeclareSIUnit\bara{bar}
\DeclareSIUnit\Curie{Ci}
\DeclareSIUnit{\msun}{\mbox{M$_\odot$}}
\DeclareSIUnit\mK{\milli\kelvin}
\DeclareSIUnit\micron{\ensuremath{\upmu}\mathrm{m}}
\DeclareSIUnit\liveday{\mbox{live-days}}
\DeclareSIUnit\tonneday{\mbox{tonne$\cdot$day}}
\DeclareSIUnit\days{\mbox{days}}
\DeclareSIUnit{\uBq}{\ensuremath{\upmu}\mathrm{Bq}}

\newcommand{\alphan}{\mbox{($\upalpha,\mathrm{n}$)}}
\newcommand{\alphanp}{\mbox{($\upalpha,\mathrm{np}$)}}
\newcommand{\alphanx}{\mbox{($\upalpha,x\mathrm{n}$)}}
\newcommand{\alphang}{\mbox{($\upalpha,\mathrm{n}\upgamma$)}}

\newcommand{\ngamma}{\mbox{($\mathrm{n},\upgamma$)}}

\newcommand{\gr}{\mbox{$\upgamma$-ray}}
\newcommand{\grs}{\mbox{$\upgamma$-rays}}

\usepackage{tabularx}

\usepackage{tabularray}
\UseTblrLibrary{booktabs}
\UseTblrLibrary{siunitx}
\usepackage{chngpage}

\newcommand{\GeantF}{\mbox{\tt Geant4}}
\newcommand{\NeuCBOT}{\mbox{\tt NeuCBOT}}
\newcommand{\NeuCBOTv}[1]{\mbox{\tt NeuCBOT-V#1}}
\newcommand{\SOURCESF}{\mbox{\tt SOURCES4}}
\newcommand{\SOURCESFA}{\mbox{\tt SOURCES4A}}
\newcommand{\SOURCESFC}{\mbox{\tt SOURCES4C}}
\newcommand{\SaGFN}{\mbox{\tt SaG4n}}
\newcommand{\TALYS}{\mbox{\tt TALYS}}
\newcommand{\TALYSv}[1]{\mbox{\tt TALYS~#1}}
\newcommand{\EMPIRE}{\mbox{\tt EMPIRE}}
\newcommand{\EMPIREv}[1]{\mbox{\tt EMPIRE~#1}}
\newcommand{\SRIM}{\mbox{\tt SRIM}}
\newcommand{\ASTAR}{\mbox{\tt ASTAR}}

\newcommand{\JENDL}{\mbox{\tt JENDL}}
\newcommand{\JENDLv}[1]{\mbox{\texttt{JENDL-#1}}}
\newcommand{\JENDLan}{\mbox{\tt JENDL/AN-2005}}
\newcommand{\TENDL}{\mbox{\tt TENDL}}
\newcommand{\TENDLv}[1]{\mbox{\tt TENDL-#1}}
\newcommand{\ENDF}{\mbox{\tt ENDF}}
\newcommand{\ENDFv}[1]{\mbox{\tt ENDF-#1}}
\newcommand{\EXFOR}{\mbox{\tt EXFOR}}
\newcommand{\CSISRS}{\mbox{\tt CSISRS}}
\newcommand{\mEXIFON}{\mbox{\texttt{mEXIFON}}}
\newcommand{\EGNASH}{\mbox{\tt EGNASH-2}}
\newcommand{\ENSDF}{\mbox{\texttt{ENSDF}}}
\newcommand{\RIPL}{\mbox{\tt RIPL}}
\newcommand{\RIPLv}[1]{\mbox{\tt RIPL-#1}}
\newcommand{\TASMAN}{\mbox{\tt TASMAN}}

\makeatletter
\renewcommand{\subsubsection}[1]{%
    \vspace{1em} 
    \begin{center} 
        \textit{#1} 
    \end{center}
    \@afterheading 
}
\makeatother

%% file: authors.tex
\author{D. Cano-Ott}\affiliation{CIEMAT, Centro de Investigaciones Energéticas, Medioambientales y Tecnológicas, Madrid 28040, Spain}
\author{S. Cebrián}\affiliation{CAPA, Centro de Astropartículas y Física de Altas Energías, Universidad de Zaragoza, Zaragoza 50009, Spain}
\author{P. Dimitriou}\affiliation{Nuclear Data Section, Division of Physical and Chemical Sciences, Department of Nuclear Sciences and Applications, International Atomic Energy Agency, POB 100, Vienna, Austria}
\author{M. Gromov}\affiliation{Skobeltsyn Institute of Nuclear Physics, Lomonosov Moscow State University, Moscow 119234, Russia} \affiliation{Joint Institute for Nuclear Research, Dubna 141980, Russia}
\author{M. Harańczyk}\affiliation{M. Smoluchowski Institute of Physics, Jagiellonian University, 30-348 Krakow, Poland}
\author{A. Kish}\affiliation{Fermi National Accelerator Laboratory, Batavia, IL 60510, U.S.A}
\author{H. Kluck}\affiliation{Institut für Hochenergiephysik der Österreichischen Akademie der Wissenschaften, 1050 Wien, Austria}
\author{V. A. Kudryavtsev}\affiliation{Department of Physics and Astronomy, University of Sheffield, Sheffield S3 7RH, UK}
\author{I. Lazanu}\affiliation{Faculty of Physics, University of Bucharest, POBox 11, 077125, Magurele, Romania}
\author{V. Lozza}\affiliation{Laboratório de Instrumentação e Física Experimental de Partículas (LIP), 1649-003, Lisboa, Portugal} 
\affiliation{Universidade de Lisboa, Faculdade de Ciências (FCUL), Departamento de Física, 1749-016 Lisboa, Portugal}
\author{G. Luzón}\affiliation{CAPA, Centro de Astropartículas y Física de Altas Energías, Universidad de Zaragoza, Zaragoza 50009, Spain}
\author{E. Mendoza}\affiliation{CIEMAT, Centro de Investigaciones Energéticas, Medioambientales y Tecnológicas, Madrid 28040, Spain}
\author{M. Parvu}\affiliation{Faculty of Physics, University of Bucharest, POBox 11, 077125, Magurele, Romania}
\author{V. Pesudo}\affiliation{CIEMAT, Centro de Investigaciones Energéticas, Medioambientales y Tecnológicas, Madrid 28040, Spain}
\author{A. Pocar}\affiliation{Amherst Center for Fundamental Interactions and Physics Department, University of Massachusetts,
Amherst, MA 01003, USA}
\author{R. Santorelli}\email{alphan@ciemat.es}\affiliation{CIEMAT, Centro de Investigaciones Energéticas, Medioambientales y Tecnológicas, Madrid 28040, Spain}
\author{M. Selvi}\affiliation{INFN - Sezione di Bologna, Bologna 40126, Italy}
\author{S. Westerdale}\affiliation{Department of Physics and Astronomy, University of California, Riverside, CA 92507, USA}
\author{G. Zuzel}\affiliation{M. Smoluchowski Institute of Physics, Jagiellonian University, 30-348 Krakow, Poland}

%% file: TexFiles/Introduction.tex
\section{Introduction}
\label{sec:Intro}

In recent years, a detailed understanding of the \alphan\ reaction has become increasingly important in various fields of physics\footnote{This paper originated from the first workshop  ``\alphan\ yield in low-background experiments'', which took place in Madrid in November 2019. Materials and conclusions from the meeting are available at this link: \href{https://agenda.ciemat.es/e/wan}{https://agenda.ciemat.es/e/wan}.}. 

Neutrons are highly penetrating particles and can produce signals that are indistinguishable from those expected in dark matter search experiments such as DarkSide~\cite{DarkSide-20k:2017zyg}, CRESST~\cite{CRESST:2019jnq}, LZ~\cite{LZ:2022ufs}, and XENON~\cite{XENON:2023sxq}. The \alphan\ reaction occurring in detector materials or surrounding structures is one of the primary sources of neutron production and represents a significant contribution to the radiogenic background in these experiments. Consequently, accurately estimating \alphan\ neutron production rates, energy spectra, and correlated $\gamma$-rays is crucial.

Neutron fluxes are computed using material assay results and computational tools that combine stopping power calculations with \alphan\ cross-sections, obtained from  measurements or theoretical models, or a combination of both. However, the accuracy of radiogenic neutron background predictions in large detectors is limited by significant uncertainties in \alphan\ yields~\cite{cooley_input_2018,kudryavtsev_neutron_2020,pigni_uncertainty_2016}. For many nuclides and materials relevant to rare event search experiments, the uncertainty in \alphan\ yield is typically within \SI{30}{\percent}, but it can reach $\mathcal{O}(\SI{100}{\percent})$ in some cases. 
The main sources of uncertainty are missing cross-section measurements, particularly for branching ratios to excited final states, or the uncertainty inherited from the theoretical models used to evaluate the \alphan\ reactions. Several mid-Z materials commonly used in low-background experiments also lack experimental data. In some cases, even when multiple experimental results are available for a specific material, discrepancies among available cross-section datasets have been observed. These discrepancies may arise from differences in the experimental setups used for the measurements, or from the corrections applied when interpreting the results to extrapolate to the total cross-section.

Furthermore, \alphan-correlated $\upgamma$-ray emission and $\upgamma$-rays from neutron capture \ngamma\ reactions extend up to \SI{10}{\MeV}, covering the energy range of interest for rare event search experiments and playing an important role in understanding detector triggers and vetoes in such experiments.

In the low-energy sector of neutrino experiments such as DUNE~\cite{DUNE:2016hlj}, \alphan\ neutrons produced by material contamination can be a potential source of background for supernova and solar neutrino studies. In particular, $\upgamma$-rays emitted in neutron capture reactions can contribute to backgrounds for neutrino-electron scattering and for neutrino-nucleus interactions such as neutrino absorption. In neutrinoless double beta decay experiments like nEXO~\cite{nEXO:2017nam}, neutrons can be captured on \ce{^136Xe} to form \ce{^137Xe}, which subsequently beta decays with a $Q$-value above the 0$\upnu\upbeta\upbeta$ energy, creating background events in the energy region of interest. The \alphan\ reactions can also create backgrounds for JUNO~\cite{JUNO:2021kxb} and SNO+~\cite{SNO:2023trz} searches, since the combination of the prompt neutron signal with the delayed capture can mimic the inverse beta decay of antineutrinos on protons. Additionally, high-energy $\upgamma$-rays can fall in the energy region of interest for nucleon decay and neutrinoless double-beta decay searches.

Understanding \alphan\ yields is also crucial in  nuclear physics, nuclear astrophysics, and certain nuclear energy-related applications. For instance, they play an essential role in the sources of neutrons for the slow neutron capture processes, in radionuclide production by energetic solar particles, in the production of positron emitters, in the nucleosynthesis of light r-process nuclei in neutrino-driven winds, and in the production of neutrons by $\upalpha$-emitters present in high-level nuclear waste.

Overall, there is a growing recognition of the need for an in-depth understanding of the \alphan\ reaction. The increasing sensitivity requirements of next-generation rare event search experiments and nuclear astrophysics have driven interest in improving \alphan\ yield calculations and assessing neutron production uncertainties.
In addition, novel techniques for measuring \alphan\ cross-sections for a variety of materials are being developed.
As such, improving the accuracy of \alphan\ yield calculations and developing new techniques for measuring \alphan\ cross-sections will gain increasing relevance as key areas of research in the coming years.

This review aims to systematically discuss the most relevant aspects of \alphan\ neutron yield calculations in low-background experiments. It compares the existing \alphan\ reaction codes and defines a common approach to uncertainties with a consistent treatment of model parameters. Moreover, it highlights the need for a more structured and accessible repository of cross-sections to facilitate their use by different codes.  Plans for measuring \alphan\ reactions relevant to underground experiments will also be presented and discussed.

%% file: TexFiles/Process.tex
\section{Process Description}
\label{sec:Porcess}

At the relatively low energy radiogenic regime (E$_\upalpha < \SI{10}{\MeV}$), the \alphan\ reaction can be described as a two-step process: 

\begin{equation}
    \mathrm{^A_ZX + ^4_2He \rightarrow ^{A+4}_{Z+2}Y^* \rightarrow ^{A+3}_{Z+2}Y + n}
\end{equation}

In the first step, the capture of the $\upalpha$ particle by the target nucleus $X$ populates an excited state of the compound $X + \upalpha \rightarrow Y^*$ nucleus. In a classical approach, the energy has to be enough to overcome the Coulomb repulsion of the two nuclei. Even though tunneling allows for the production of classically forbidden compound nuclei for a certain available energy, this is an important parameter percolating in the cross-section for this reaction. As a consequence, the nuclei expected to contribute more to neutron production are those with low-$Z$, given the small Coulomb barrier. This notwithstanding, some mid-$Z$ metals might have sizable contributions due to specific nuclear properties \cite{Wilson2009}. 

The second step of the \alphan\ reaction is where the parameter space can be diverse. In the simplest scenario, the neutron drips off the compound nucleus $^{A+4}Y$. This is allowed when the populated $^{A+4}Y^*$ state is above a certain threshold (denoted as $S_\mathrm{n}$). Assuming that no other particles are involved and the decay proceeds to the ground state of $^{A+3}Y$, the neutron energy can be expressed as:

\begin{equation}
 E_\mathrm{n} = \frac{A+3}{A+4}[E(Y^*) + (M(^{A+4}Y)-M(^{A+3}Y))c^2],
\end{equation}

where $E$ denotes  energy of a state or a particle in the center of mass frame or a state and $M$ the mass of a nucleus, since all the states above $S_\mathrm{n}$ are in the continuum, $E$ can have any value above

\begin{equation}
E_\mathrm{n} = \frac{A+3}{A+4}[S_\mathrm{n} + (M(^{A+4}Y)-M(^{A+3}Y))c^2].
\end{equation}

Beyond the simplest case of neutron emission, the reaction mechanism can involve more complex decay pathways. The state of $^{A+3}Y$ following neutron emission might not be the ground state, resulting in an \alphang\ reaction or a multiple particle decay. 
How this complex behavior is typically modeled in a statistical approach is summarized in Section~\ref{sec:Xsec:Models}, and the associated uncertainties are discussed in Section~\ref{sec:Uncertainty:xsec}.

Detecting n-\(\upgamma\) coincidences and neutron emission with multiplicity of 2 or larger is challenging. Both \(\upgamma\)-rays and neutrons have relatively large mean free paths compared to other types of ionizing radiation. This is particularly true for neutrons, which are inherently difficult to detect due to their low interaction probability. Consequently, neutron coincidences are often characterized by extremely low detection efficiency. Achieving high-resolution and statistically significant measurements of such final states necessitates the use of detectors with extensive coverage, high granularity, and experiments utilizing high-intensity beams (see for instance \cite{Martinez:2014}). As a result, data for \alphang\, \alphanp\ and {($\upalpha,X\mathrm{n}$)} cross-sections—where $X$ denotes any number of additional neutrons—are particularly limited. The term \alphan\ is often used inclusively, neglecting this richness, to refer to the entire family of \alphanx\ processes, where $x$ represents any additional particle emitted alongside a single neutron (see Section IV for a discussion on the available data and calculations).

The overall contribution of these higher multiplicity channels is, however, subdominant. Where ($\mathrm{\upalpha, 2n}$) is most energetically favorable among the light elements are $\mathrm{^{18}}$O (7.5 MeV), $\mathrm{^{22}}$Ne (\SI{7.8}{\MeV}) and $\mathrm{^{26}}$Mg (\SI{8.4}{\MeV}), which are barely accessible with radiogenic $\mathrm{\alpha}$-particles --- this channel is only open for the \SI{8.95}{\MeV} $\mathrm{\alpha}$ decay of $\mathrm{^{212}}$Po in the $\mathrm{^{232}}$Th chain --- and higher multiplicities are even more disfavored. 

%% file: TexFiles/Isotopes.tex
\section{Importance of \texorpdfstring{\alphan}{(alpha,n)} reactions in different fields and relevant nuclides}

\label{sec:Isotopes}

\subsection{Searches for rare events}
\label{sec:Inputs:Isotopes:LowBG}

Neutron radiation from \alphan\ reactions is a serious concern for rare event search experiments that require very low levels of radioactive background, such as direct detection of dark matter or searches for neutrinoless double beta decay.
If the detector's components or materials are contaminated with naturally occurring radioactive nuclides, e.g. \ce{^235U}, \ce{^238U} and \ce{^232Th}, their  decay chains, which contain many $\upalpha$-emitters, can produce a considerable neutron background that would limit experimental sensitivity. Thus, the material composition should always be carefully selected and used as an input to Monte Carlo simulations that are performed to evaluate potential backgrounds, as explained in Section~\ref{sec:Tools}.

\subsubsection{\alphan\ induced backgrounds}
\label{sec:Inputs:Isotopes:LowBG:alpha-n}

A material often used in low-background experiments because of its high radio-purity is polytetrafluoroethylene (PTFE, or Teflon\textsuperscript{TM}). PTFE also features good dielectric properties and high reflectivity for vacuum-ultraviolet (VUV) light \cite{LZ:2022ufs,Haefner:2016ncn}. It is commonly used in cryogenic environment, e.g., for a variety of insulating structural elements, including cables supports. It is also used to reduce friction between metal parts, and as a reflector to enhance collection of light generated by plastic/liquid scintillators and noble elements, such as argon and xenon. 

PTFE contains fluorine whose only stable isotope is \ce{^19F}, which has a low threshold (around \SI{2.3}{\MeV}) for \alphan\ reaction and a steeply rising cross-section with energy. While several measurements and model predictions of this cross-section exist, experimental data exhibit significant variation, and the resonance-like behavior is not well described by these models (see also Fig.~\ref{fig:figF} in Sec.~\ref{sec:Xsec}).

Other nuclides with \alphan\ cross-sections relevant for rare event searches include: 
\begin{itemize}  \setlength\itemsep{0em}
    \item carbon (\ce{^13C}) contained in plastics, polyethylene, PTFE, scintillators and e.g. in rock surrounding underground experimental caverns,
    \item nitrogen (\ce{^14N}) present in some plastics and often used as inert buffer gas,
    \item oxygen (\ce{^17O}, \ce{^18O}) abundant in plastics, quartz, rock and water,
    \item silicon (predominantly \ce{^29Si} and \ce{^30Si}) contained in various types of glass and quartz and used for widely used semiconductor detectors,
    \item aluminium (\ce{^27Al}), present in ceramics and sapphire,
    \item titanium, (stainless) steel, and copper (all naturally occurring isotopes), that are used to build cryostat vessels, support structures, and shielding, 
    \item beryllium (\ce{^9Be}) present in wires and connection pins made out of CuBe alloy,
    \item sodium, chlorine, calcium and other elements found in rocks.
\end{itemize}


Considering natural radioactivity, besides the \alphan\ reactions, spontaneous fission also contributes to neutron production:
Spontaneous fission (SF), as described by Watt's formulae~\cite{watt}, produces the same neutron yield for all materials and depends only on the concentration of the fissioning nuclide. Among naturally occurring radioactive nuclides, only fission of \ce{^238U} contributes significantly to neutron production. Although the probability of SF of \ce{^238U} is about \num{5e-7} compared to the $\upalpha$ decay rate, the neutron yield from this process dominates over that from \alphan\ reactions for high-$Z$ materials where the neutron production is highly suppressed due to the Coulomb barrier. In practice, neutrons produced in the SF process can be tagged with high efficiency in a detector or a veto system due to simultaneous emission of several neutrons and \grs. 

The neutron yield from \alphan\ reactions depends on the energy of $\upalpha$-particles, their energy loss in a particular material, and the cross-section of the \alphan\ reaction. The two most common natural radioactive decay chains, \ce{^232Th} and \ce{^238U}, are critical in the calculation of neutron background for low-background experiments. They produce 6 and 8 $\upalpha$ decays, respectively. The energies of each $\upalpha$ in the decay chain need to be considered. The decay chain of \ce{^235U}, which comprises 7 $\upalpha$ decays, also contributes to neutron production,  although to a lesser extent due to the small abundance of \ce{^235U} in natural uranium (\SI{0.72}{\percent}). Finally, secular equilibrium in the decay chains is often broken, especially for \ce{^238U}.

\begin{center}
\subsubsection{Material screening techniques to mitigate $\upalpha$ emission}
\label{sec:Inputs:Isotopes:LowBG:screening}
\end{center}

\begin{figure}
\centering
\includegraphics[width=0.8\textwidth]{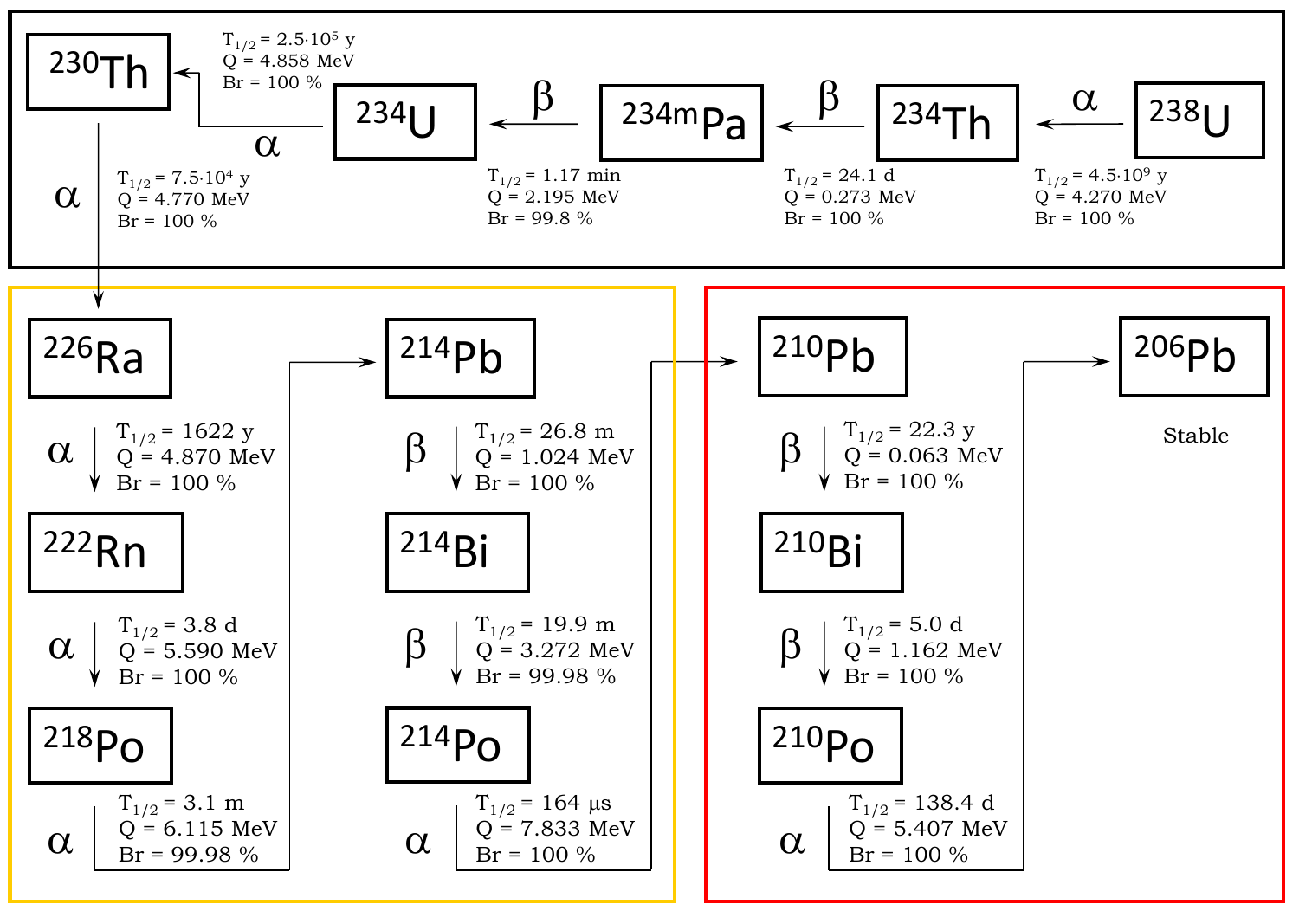}
\caption{\label{fig:U_chain} The \ce{^238U} decay chain is divided into three parts, in which the long-lived uranium/thorium isotopes (top), \ce{^226Ra} with \ce{^222Rn} and its short-lived daughters (bottom left), and the long-lived \ce{^222Rn}~daughters (bottom right) appear. The three parts are usually in disequilibrium thus, they should be assayed separately.}   
\end{figure}

 For illustration, in Fig. \ref{fig:U_chain} a simplified decay scheme of the \ce{^238U} decay chain is shown. The black box includes the top part of the chain with the long-lived uranium/thorium isotopes (3 $\upalpha$-decays). The yellow box indicates the middle part of the chain with \ce{^226Ra}, \ce{^222Rn} and its short-lived daughters (4 $\upalpha$-decays). Finally, the bottom part of the chain is shown in the red box, including the so-called ``long-lived \ce{^222Rn} daughters" (one $\upalpha$-decay). The secular equilibrium is usually broken at the level of \ce{^226Ra} and \ce{^210Pb}. Due to the relatively short half-lives of their daughters, we can assume that the respective sub-chains are in equilibrium. Because radium is an alkaline earth metal, while uranium and thorium are actinoids, chemical processes that affect alkaline earth metals—such as solubility in water and ion exchange—can lead to radium enrichment or depletion in materials undergoing production. Due to radium's long half-life, these changes can persist for a significant time before equilibrium is re-established. Moreover, due to the \SI{22}{\yr} half-life of \ce{^210Pb} there is usually disequilibrium between the short- and the long-lived \ce{^222Rn} daughters. This is why it is not recommended to conclude about the specific activities of \ce{^226Ra} (and \ce{^222Rn}) from e.g. the ICP-MS measurements (determination of uranium/thorium). The same concerns the bottom part of the \ce{^238U} chain: it is usually not possible to predict the activities of \ce{^210Pb}--\ce{^210Po} from the high-sensitivity \ce{^222Rn} emanation measurements, or from the $\upgamma$-ray screening. Therefore, in order to predict properly the background rates caused by a particular element of the detector, it is necessary to assay each part of the chain separately. Clearly, depending on the experiment and its goal, different sub-chains may contribute differently, but usually the most important is \ce{^226Ra}/\ce{^222Rn} and their long-lived decay products. Therefore, in the last years strong emphasis has been put on developments of high-sensitivity $\upgamma$-ray spectrometers, radon emanation techniques and $\upalpha$-spectroscopy.

A disequilibrium effect has been observed for example for the nylon foil, which was used to construct the Borexino scintillator vessel \cite{Zuzel:2003}. In most of the investigated samples the \ce{^226Ra} activity concentrations were higher compared to \ce{^238U}, but the last one showed a different behavior: radium content was a factor 20 lower compared to uranium and the effect was related to a different production process of the foil.

The foundation of the experiments devoted to searches for rare events is an extensive radioassay program. Its goal is to perform a selection of materials and components, which can be used for construction of respective detectors. The selection is based on the content of radioactive nuclides, where the allowed concentrations are determined applying extensive Monte Carlo simulations. 
Determined residual radioactivity for various components can vary significantly from different suppliers and even from production batches. A great deal of radio-purity measurements are available for materials like copper, stainless steel, PTFE, polyethylene, and for electronics and readout components (see for instance results collected in the radiopurity database\footnote{Available at: \url{http://radiopurity.org}.} and in Refs. \cite{LEONARD2008490,Cebrian_2015,ABGRALL201622,LEONARD2017169,Cebrian_2017,XENONrpurity,LZrpurity,GERDArpurity}). An accurate estimate of \alphan\ neutron yields requires the knowledge of radioactivity in the material to normalize the MC simulations.

Different techniques are commonly used in material screening. Mass spectrometry, like Inductively Coupled Plasma Mass Spectrometry (ICP-MS) or Glow Discharge Mass Spectrometry (GDMS) are used to detect uranium and thorium in samples. Thus, only the activity concentrations of nuclides in the upper part of the chains of \ce{^238U} (upper box in Fig.~\ref{fig:U_chain}) and \ce{^232Th} can be deduced. However, mass spectrometers are usually very sensitive and allow to perform measurements down to a sub-ppt (parts per trillion, \num{e-12} g/g) level~\cite{LaFerriere:2015}. 

The $\upgamma$-ray counting is a powerful and sensitive method to look for radio-impurities, without destroying the sample. It allows investigating the middle part of the \ce{^238U} (mainly \ce{^226Ra} and its daughters - see the yellow box in Fig.~\ref{fig:U_chain}) and \ce{^232Th} chains, as well as to \ce{^40K}, \ce{^60Co}, \ce{^137Cs} and other $\upgamma$-emitters. Assay time depends on the required sensitivity and can be as long as a couple of months. Sample size is usually limited to a few tens of kilograms because of the effect of self-shielding. There are many $\upgamma$-counting facilities around the world, but the most sensitive instruments are operated at the INFN Laboratori Nazionali del Gran Sasso and at the Canfranc Underground Laboratory reaching sensitivities of \SI{10}{\uBq\per\kg}~\cite{Neder:2000, Zuzel:2023}. In some cases (e.g. thin foils) \ce{^226Ra} may be investigated by application of high sensitivity \ce{^222Rn} emanation techniques, reaching similar or even better sensitivities~\cite{zuzel:2017} compared to $\upgamma$-ray spectrometry.

To assay \ce{^210Pb} in the bottom part of the \ce{^238U}-chain (red box in Fig.~\ref{fig:U_chain}) chemical extraction of \ce{^210Po} is applied. The advantage of this method is that usually only a small mass is needed (a few g) for analysis. High purity concentrated acids like \ce{HCl}, \ce{HNO_3}, \ce{HF}, \ce{H_2SO_4} are used to dissolve samples. They are always spiked with \ce{^208Po} or \ce{^209Po} for determination of the chemical yield (Po extraction efficiency). If organic matter is present, also \ce{H_2O_2} is added. Next, the mixture is converted into a \SI{0.5}{\molar} \ce{HCl} solution, from which polonium is auto-deposited on a silver disc. In some cases, separation of Po from the matrix on a dedicated column is necessary. The activity of polonium is measured with a low background $\upalpha$-spectrometer, with the overall efficiency deduced from the spike signal. Sensitivities down to \SI{0.5}{\milli\becquerel\per\kg} for \ce{^210Po} are achievable. As an example, it has been found that the \ce{^210Po} concentrations in copper follows the chemical purity of the material and changes from \SI{10}{\milli\becquerel\per\kg} (oxygen-free copper) up to \SI{10}{\becquerel\per\kg} (fire-refined copper). Values obtained for low radioactivity lead (\SI{\approx 2}{\becquerel\per\kg}) were consistent with \ce{^210Pb} measured via beta spectroscopy~\cite{zuzel:2012}. Activity concentrations deduced for high-purity titanium were at the level of \SI{2}{\becquerel\per\kg}. The results confirmed the first assays of \ce{^210Po} bulk activities in metals performed with application of large surface low background $\upalpha$-spectrometer~\cite{zuzel:2017_2}. 

Another class of background source is the surface contamination with naturally occurring $\upalpha$-emitters (mostly the short- and long-lived daughters of \ce{^222Rn}). Exposure to environmental radon during fabrication, assembly, and installation of low background systems can lead to build-up of \ce{^210Pb} on surfaces being in contact with the active detector parts. \ce{^210Pb} in the sub-surface layer can be accumulated though the diffusion of radon. \ce{^222Rn} atoms can penetrate layers up to several tens of \SI{}{\micron}
depending on the diffusion constant~\cite{wojcik:2000}. \ce{^210Pb} which has a \SI{22}{\yr} half-life will therefore act as an approximately constant source of radiation (from self decays and from decays of \ce{^210Bi} and \ce{^210Po}) throughout the full life of an experiment. In case of surface contamination, the $\upalpha$-decays are relatively easy to identify by their specific energies (e.g. \SI{5.3}{\MeV} for \ce{^210Po}). If the $\upalpha$-peaks are shifted towards lower energies due to quenching effects (e.g. in the scintillators) the decays can still be recognized by application of the pulse shape analysis. Decays of $\upalpha$-emitting nuclides occurring on the surfaces of materials with high \alphan\ cross-sections (e.g., PTFE) will significantly contribute to the neutron background. Sensitive surface and bulk assay of \ce{^210Po} may be carried out using low background large surface $\upalpha$-spectrometers, like the XIA UltraLo-1800~\cite{zuzel:2017_2}. Due to very low background and by analysis of the continuous part of the spectrum between \SI{1.5}{\MeV} and \SI{6.0}{\MeV} bulk \ce{^210Po} may be assayed down to about \SI{50}{\milli\becquerel\per\kg}~\cite{Zuzel:2003}. Looking at the peak around \SI{5.3}{\MeV} surface contamination as low as \SI{0.5}{\milli\becquerel\per\square\m} may be determined~\cite{zuzel:2017_2}. To avoid problems related to surface contamination, it is recommended to store all detector components after production and before installation in a radon-free atmosphere (in clean rooms supplied with radon-free air or in multi-layer plastic bags that are impervious to radon). Also, material-specific surface cleaning protocols should be applied.

Particular attention should also be paid to the final machining of structural materials. Modern detectors devoted to searches for rare processes are quite large and include tonnes of materials, such as metals or plastics. A small radiogenic contamination per unit mass/surface can lead to a significant \alphan\ background. Taking into account that the material is usually purchased not in the form of final products, but in the form of intermediate goods based on the results of the assays, extra amount of uranium and thorium may occasionally be introduced into structural components of the detector during the manufacturing and later machining processes. Therefore, the assays should be repeated after every stage of production, if possible. Ideally, such constant monitoring of radioactive contamination should be included in the technological chain of production of a structural element of the detector from the stage of purchasing raw materials to the stage of manufacturing the final product. Fortunately, there are several technologies that are already available and applied. This refers to the production of pure PMMA with/without gadolinium \cite{Gd-PMMA-2021} and titanium \cite{Mozhevitina:2015jla, Chepurnov:2013boc}.

\vspace{1cm}
\subsection{Nuclear astrophysics}
\label{sec:Inputs:Isotopes:NuclearAstrophysics}


The nucleosynthesis, or creation of the elements in the Universe, is carried out via nuclear reactions \cite{bothe1939, RevModPhys.29.547}. Charged particle reactions, especially those induced by the primary products of Big Bang nucleosynthesis (protons, \ce{^2H}, \ce{^3He}, and $\upalpha$-particles) are the main process in the synthesis of the elements with $A<60$. The overlap between the Coulomb barrier penetration probability and the temperature distribution of particle energies will determine the probability of these reactions to take place \cite{Krane:359790,10.3389/fspas.2020.00009}.

The fundamental reaction in stars is the proton-proton fusion. After the hydrogen fuel is depleted, the star undergoes gravitational collapse reaching a higher temperature (\SI{\approx e8}{\kelvin}), thus the Coulomb barrier for \ce{^4He}-\ce{^4He} fusion can be overcome \cite{Krane:359790}.

The \alphan\ reactions are crucial in various astrophysical mass ranges, providing insights into the origin of the elements. They are particularly important in the weak r-process (also called the $\alpha$-process), which occurs in the neutrino-driven ejecta of core-collapse supernovae and is responsible for producing lighter heavy elements observed in metal-poor stars. The uncertainty in their astrophysical rates is the primary factor affecting this process. In stars, the slow neutron-capture process (s-process) is one of the two main mechanisms responsible for the formation of elements heavier than iron. The efficiency of this process is heavily dependent on \alphan\ reactions, which serve as the primary sources of neutrons that initiate the neutron-capture chain, resulting in the creation of elements up to bismuth.

The \ce{^22Ne}\alphan\ce{^25Mg} and \ce{^13C}\alphan\ce{^16O} reactions are the main sources of neutrons for the s and i-processes (intermediate neutron capture process \cite{2014nic..confE.145D,Hampel_2016,Hampel_2019}), while \alphan\ reactions on $17<A<34$ nuclei may impact nucleosynthesis in Type Ia supernova explosions. In the case of \ce{^22Na}, the cross-section is mostly unknown in the relevant stellar energy range of \SIrange{450}{750}{\keV}, where direct measurements provide only upper limits, and estimates from indirect sources are uncertain.
\ce{^13C}\alphan\ce{^16O} reaction has been studied in \cite{PhysRev.107.1065,PhysRev.156.1187, DAVIDS1968619,PhysRevC.7.1356,1993ApJ...414..735D,Kellogg1989,PhysRevC.72.062801,PhysRevC.105.024612,PhysRevLett.129.132701} and experimental cross-sections exist in \SIrange{0.3}{8}{\MeV} energy range.

In the low-mass regime, the \ce{^9Be}\alphan\ce{^12C} reaction is critical for both the s and r-processes, as well as for primordial nucleosynthesis.

For core-collapse supernovae explosions, particularly important reactions include \ce{^96Zr}\alphan\ce{^99Mo}, \ce{^100Mo}\alphan\ce{^103Ru}, \ce{^86Kr}\alphan\ce{^99Sr}.

The cross-sections of the \ce{^96Zr}\alphan\ce{^99Mo} and \ce{^100Mo}\alphan\ce{^103Ru} reactions were recently measured at energy ranges relevant to astrophysics \cite{Kiss2021, PhysRevC.104.035804}. These measurements were conducted to better understand the parameters of the $\upalpha$+nucleus optical potential, for the latter see also Section \ref{sec:Xsec:Models}.

Recently, the \ce{^86Kr}\alphan\ce{^89Sr} cross-section at an energy relevant for the weak r-process in the neutrino-driven winds of core-collapse supernovae has been measured in order to reduce uncertainty in model predictions for nucleosynthesis in this site \cite{refId0}.

\subsection{Neutron sources}
\label{sec:Inputs:Isotopes:AlphaNeutronSources}


The combination of an $\upalpha$ emitter and a low-$Z$ material is a typical method exploited to produce inexpensive and compact neutron sources, such as those used for calibration purposes in various types of experiments. Beryllium is a very common material used in $\upalpha$-neutron sources, which can yield approximately \num{e-4} neutrons per $\upalpha$-decay through the reaction:

\begin{equation}
\upalpha + \ce{^{9}Be} \rightarrow \ce{^{12}C} + \mathrm{n} + \upgamma \quad \text{(\SI{4.44}{\MeV}).}
\label{eq:be(a,n)cg}
\end{equation}

Fluorine, lithium, carbon, and boron are other alternative materials with high $\upalpha$-neutron cross-sections.

As for typical $\upalpha$-emitters, the list includes \ce{^{241}Am}, \ce{^{238}Pu}, \ce{^{239}Pu}, \ce{^{210}Po}, and \ce{^{226}Ra} (see, for instance, \cite{Gd-PMMA-2021}). The strength of the neutron source is determined by the activity of the $\upalpha$ emitter. Activities in the range of approximately \SIrange{1}{1000}{\giga\becquerel} are common, although smaller or larger values are possible for special purposes. The actual neutron yield depends on the source matrix and, in particular, on the concentration of the $\upalpha$-emitting nuclide within the matrix.

The average (maximum) neutron energy of Am-Be and Pu-Be sources is \SI{\approx4.2}{\MeV} (\SI{11}{\MeV}). For Ra-Be source it is \SI{3.6}{\MeV} (\SI{13.2}{\MeV}). 
Am-F and Am-Li sources produce neutrons with relatively low average energies  of \SI{1.5}{\MeV} and \SI{0.5}{\MeV}, respectively, making Am-Li, for example, a suitable calibration source for detectors intended to measure delayed neutrons, such as those used in fission experiments.

It is important to note that these reactions may involve the emission of $\upgamma$-radiation. For example, the associated 4.44~MeV $\upgamma$-emission also makes Am-Be sources suitable for gamma calibration, particularly in experiments requiring high-energy gamma-ray references.

\subsection{Nuclear technologies}
\label{sec:Inputs:Isotopes:NuclearFuel}

Materials control and accountancy of uranium and plutonium throughout the fuel cycle heavily depend on a diverse range of passive and active neutron counting methods. Given that these materials are typically found in compound forms like oxides, fluorides, and carbides, with potential light element impurities such as lithium, beryllium, and boron, the production of neutrons via \alphan\ reactions often constitutes a notable source of neutron signals. Additionally, this phenomenon contributes significantly to self-interrogation in items undergoing multiplication \cite{Favalli2019}.

Neutrons and \grs\ emitted during the \alphan\ process contribute to the total flux of radiation along the fuel cycle, from the enrichment and fuel fabrication, fuel reprocessing, and disposal of the spent fuel. More accurate data on the cross-sections, total neutron yields, neutron spectra, and \gr\ emissions from \alphan\ reactions are needed for reducing the (frequently large) uncertainties in neutron background and activation estimation, nuclear waste characterization, dosimetry, nondestructive mass assay of fresh and used nuclear fuel, nuclear safeguards, and materials control and accountancy. The thick target integrated over angle yield curve is perhaps the most important quantity for applications. However, thin-target data, neutron and \gr\ spectra are also needed, in particular data on partial cross-sections and angular distributions, for the calculation and evaluation of 4$\pi$ emission spectra. 

$^{19}$F\alphan\ and $^{17,18}$O\alphan\ can be considered as the most relevant reactions for fission reactor technologies. Other important nuclides in fission applications are lithium, beryllium, boron, carbon, nitrogen, sodium, aluminum, and silicon. The renewed interest in molten salt reactors may require new data on different salt components like lithium, beryllium, nitrogen, fluorine, sodium, and potassium. In addition, the development of fusion reactors requires improving \alphan\ data on structural materials and materials used for plasma diagnostics. 

\subsection{Medical applications}
\label{sec:Inputs:Isotopes:MedicalApplications}
 

The vast majority of radiopharmaceuticals produced worldwide are used in diagnostic applications, particularly in imaging techniques such as SPECT and PET.
These methods allow for the assessment of organ integrity and function, as well as the metabolic pathways of radio-tracers in the body.
Notwithstanding the economic aspects, to be efficient and safe to use, diagnostic radiopharmaceuticals should give a low dose of radiation, should be eliminated quickly from the body (having short effective half-lives) and should be trapped by the metabolic process of interest. Thus, a very specific and limited number of nuclides is considered. Radionuclides used for these applications can be produced in nuclear reactors or through the irradiation of a target with particle accelerators using different particle beams. 
In principle, light charged particles --- protons (p), deuterons (d), \ce{^3He}, and $\upalpha$\text{s} --- can be used for medical nuclide production.
Protons are most commonly employed due to the availability of small to medium-sized cyclotrons at hospitals, while d, \ce{^3He}, and \(\upalpha\)-particles have more limited applications compared to protons. $\upalpha$s can induce high cross-section reactions, especially in light and medium-mass target nuclei, such as \((\upalpha, x\text{n})\) reactions. However, $\upalpha$-particles have a shorter range in the target material, involving fewer target nuclei and resulting in lower yields compared to other particles. Another disadvantage of $\upalpha$-induced reactions is that the available $\upalpha$-beams have lower intensity than other charged-particle beams. Despite this, $\upalpha$-particles have three major advantages: (a) certain radionuclides can only be produced through $\upalpha$-induced reactions, (b) high-spin isomeric states are preferentially produced, and (c) the product radionuclide is often two charge units higher than the target, allowing for more specific chemical separation and high purity. Thus, $(\upalpha, x\mathrm{n})$ reactions represent an important option.

Some of the radionuclides commonly produced using $\upalpha$-particles are listed in Table \ref{tabisotopes} \cite{Mohr2015}. 
In the IAEA Medical Isotope Browser\footnote{Available at: \url{https://nds.iaea.org/mib}.}, the production channels can be selected based on the optimum energy for maximum $\upalpha$ cross-section and thick target yield, as well as the purity of the produced radionuclide. Examples of ($\alpha$, n) production routes that are used include: $^{35}$Cl($\alpha$, n)$^{38}$K (7-22 MeV) and $^{92}$Mo($\alpha$, n)$^{96}$Ru (16-28 MeV), with optimum energy windows generally above 10 MeV. Other ($\alpha$, n) reactions typically used for medical applications are: $^{63}$Cu($\alpha$, n)$^{66}$Ga (7-35 MeV) for the production of therapeutic $^{66}$Ga, and $^{40}$Ca($\alpha$, n)$^{43}$Sc (5-35 MeV) and $^{66}$Zn($\alpha$, n)$^{69}$Ge (8-35 MeV) for the production of diagnostic radionuclides $^{43}$Sc and $^{69}$Ge (see IAEA Medical Portal\footnote{Available at: \url{https://nds.iaea.org/medportal}.}).

\begin{table}[!h]
\centering
\caption{Radionuclides commonly produced using $\upalpha$-particles.}
\label{tabisotopes}
\begin{tblr}{cc} 
\toprule
Range of negative $Q$-values / \si{\MeV} & Residual nuclide produced in the \alphan\ reactions\\\midrule
 {[0,2]} & \ce{^53Mn}, \ce{^51Ti}, \ce{^22Na}  \\ 
  {(2,4]} & \ce{^54Mn}, \ce{^51Cr}, \ce{^48V}, \ce{^44Sc}, \ce{^43Sc}, \ce{^40K}, \ce{^36Ar}, \ce{^33S}, \ce{^30P}, \ce{^26Al} \\ 
 {(4,6]} & \ce{^53Fe}, \ce{^49Cr}, \ce{^45Ti}, \ce{^38K}, \ce{^37Ar}, \ce{^34Cl}  \\ 
 {(6,8]} & \ce{^42Sc}, \ce{^27Si}, \ce{^23Mg}  \\ 
 {(8,10]} & \ce{^47Cr}, \ce{^39Ca}, \ce{^35Ar}, \ce{^31S}  \\ 
$>10$ & \ce{^43Ti}, \ce{^21Mg}  \\
\bottomrule
\end{tblr}
\end{table}

Other types of nuclides considered in medical applications that can be produced by the reactions: 

\begin{enumerate}

\item
Radioactive isomeric states of a few nuclides have very suitable decay properties for therapeutic applications. In general, they are low-lying states with high nuclear spins. They decay mostly to their respective ground states by a high internal conversion transition. The low-energy conversion electrons, or an avalanche of emitted Auger electrons, can lead to precise localized internal therapy effects if the radioactive species is properly attached to an appropriate chemical carrier. Several such isomeric states can be mentioned: \ce{^{117m}Sn}, \ce{^{193m}Pt} and \ce{^{195m}Pt}, obtained in reactions $(\upalpha,x\mathrm{n})$, with $x=1$ and $3$ from \ce{^116Cd} and \ce{^192Os}~\cite{Syed2016}.

\item
Certain halogen radionuclides are widely used for both diagnosis and therapy in nuclear medicine. 
Fluorine (\ce{^18F}, typically produced via \ce{^18O}$(\text{p},\text{n})$\ce{^18F}, with a half-life of 110 minutes) forms the strongest carbon-halogen bond, but it is limited to PET imaging and has too short a half-life to study slow metabolic processes, such as those involving proteins and peptides \cite{Breunig2017}. It is worth noting that the production of \ce{^18O}$(\text{p},\text{n})$\ce{^18F} also leads to secondary neutron emission, which may pose radiation safety challenges in medical cyclotron facilities.
Iodine, on the other hand, has some isotopes obtained from \ce{^123Sb}$(\upalpha, x\mathrm{n})$\ce{^{125,124}I} reactions, which are suitable for a broader range of medical applications. However, its chemical bonding is sometimes too weak, leading to instability of labeled molecules in vivo.
Radionuclides from the bromine family are an alternative \cite{WELCHMCELVANY+1983+41+46}. The positron emitters \ce{^75Br} ($T_{1/2} = \SI{96.7}{\minute}$) and \ce{^76Br} ($T_{1/2} = \SI{16.2}{\hour}$) could be used for medical applications. The \ce{^77Br}, decaying by electron capture (half-life of \SI{57}{\hour}) is a candidate in Auger-therapy, as well as the shorter-lived \ce{^{80m}Br} ($T_{1/2} = \SI{4.4}{\hour}$), decaying by isomeric transition, while the $\upbeta^-$-emitter \ce{^82Br} ($T_{1/2} = \SI{35.2}{\hour}$) is also a candidate for therapy applications. Reactions induced by $\upalpha$-particles on arsenic \ce{^75As} lead to \ce{^{76,77,78}Br}. 	This production route is only explored using particle accelerators. The latest evaluation of \ce{$^{75}$As($\alpha$, 3n)$^{76}$Br} is in Ref. \cite{Tarkanyi2019}. 

\item
A new option of cancer treatment is possible with the radioactive nuclide \ce{^211At} \cite{doi:10.1089/cbr.2012.1292,10.3389/fmed.2022.1076210}. Results suggest that the short 7.2-hour half-life of \ce{^211At} can provide blood-borne cancer patients with just enough radiation therapy to target their cancer cells and minimizes exposure of the rest of the body; it also limits the exposure of the team that manipulated this radioisotope. The nuclear reaction that describes the production process of \ce{^211At} in a bismuth target is \ce{^209Bi}$(\upalpha,2\mathrm{n})$\ce{^211At} \cite{LARSEN1996135,Tarkanyi2022}. This production route is only explored using particle accelerators.

\item
Nuclides \ce{^43K} and \ce{^30P} are also of interest for biological and medical studies. These can be produced via the reactions \ce{^40Ar}$(\upalpha, \text{p})$\ce{^43K} and \ce{^27Al}$(\upalpha, \text{n})$\ce{^30P} \cite{Mohr2015}. Medically important nuclides also include \ce{^{110m}In} and \ce{^111In} \cite{Ditroy2018}. Typically, \ce{^111In} is obtained in a cyclotron via the reaction \ce{^112Cd}$(\text{p},2\text{n})$\ce{^111In} \cite{Tsopelas2015,Tarkanyi2019a}, or through reactions such as \ce{^109Ag}$(\upalpha,2\text{n})$\ce{^111In}.  For \ce{^{110m}In}, possible production routes include \ce{^107Ag}$(\upalpha,\text{n})$\ce{^{110m}In} and \ce{^109Ag}$(\upalpha,3\text{n})$\ce{^{110m}In}. The alternative production route \ce{^107Ag}$(\upalpha, \text{n})$\ce{^{110m}In} has been also investigated in \cite{Tarkanyi2019}.

\end{enumerate}

The activation of materials in the beam of $\upalpha$-particles, i.e. the formation of radioactive products, is an interesting topic of study both as fundamental research and for applications. The formation of nuclear isomeric states is difficult to reproduce by theory. The cross-section data obtained as a function of $\upalpha$-particle energy are of great practical significance in the production of some medical-related radionuclides, although their yields are generally much lower than in proton or deuteron-induced reactions. There are several radionuclides that are exclusively produced through the use of $\upalpha$-particles, and there are a few low-lying high-spin isomeric states that are preferentially populated in $\upalpha$-particle induced reactions. 


  
  
%

%% file: TexFiles/Data.tex
\section{Measured and calculated  \texorpdfstring{\alphan}{(alpha,n)} cross-sections}
\label{sec:Xsec}

\begin{figure}[ht]
    \centering
    \begin{minipage}{.5\textwidth}
        \includegraphics[width=0.99\linewidth]{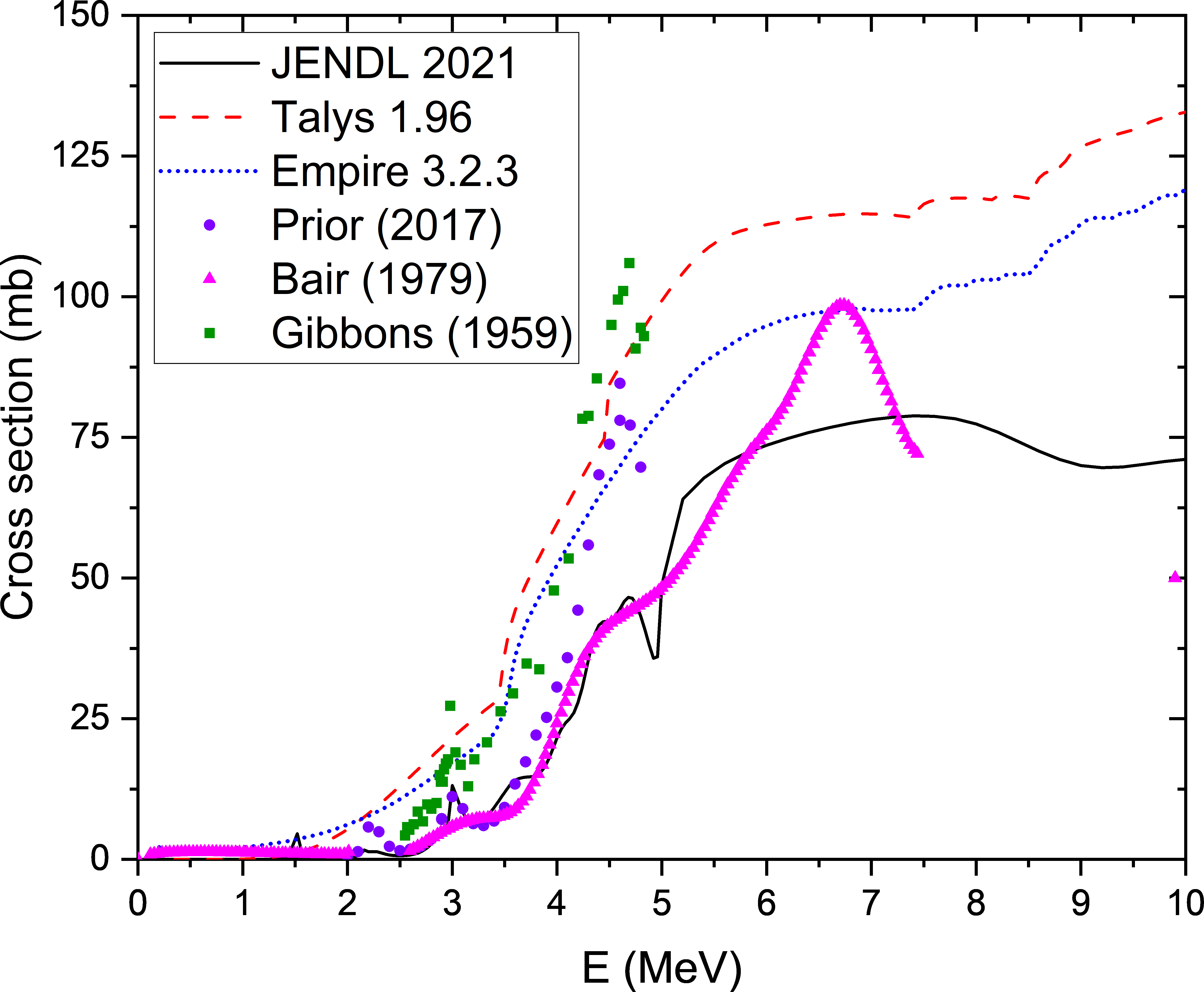}
    \end{minipage}%
    \begin{minipage}{.5\textwidth}
        \includegraphics[width=0.99\linewidth]{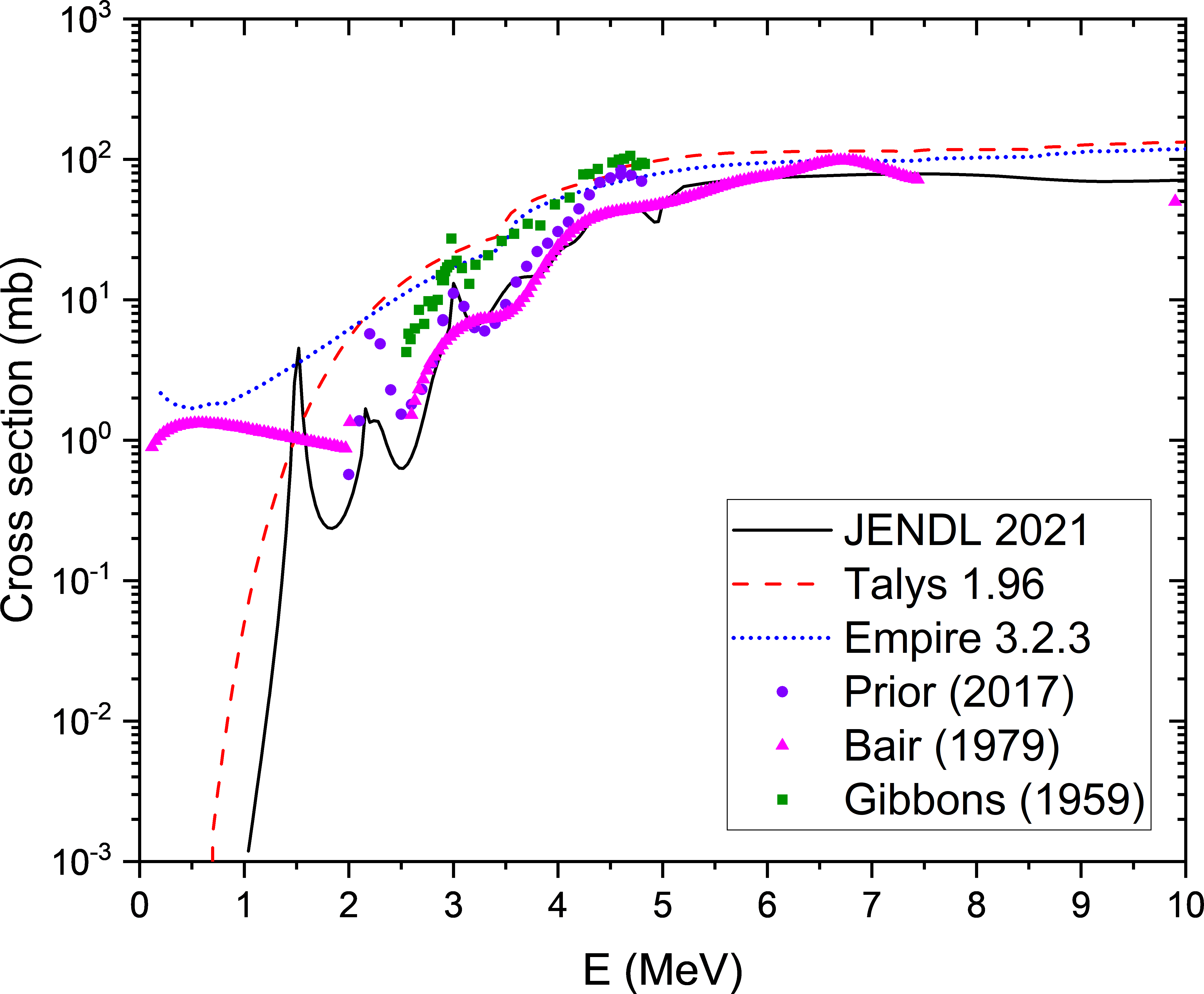}
    \end{minipage}
    \caption{\ce{^{10}B} \alphan \ce{^{13}N} cross-section as a function of energy in linear (left) and logarithmic scale (right) across different orders of magnitude.}
    \label{fig:figB10}
\end{figure}

\begin{figure}[ht]
    \centering
    \begin{minipage}{.5\textwidth}
        \includegraphics[width=0.99\linewidth]{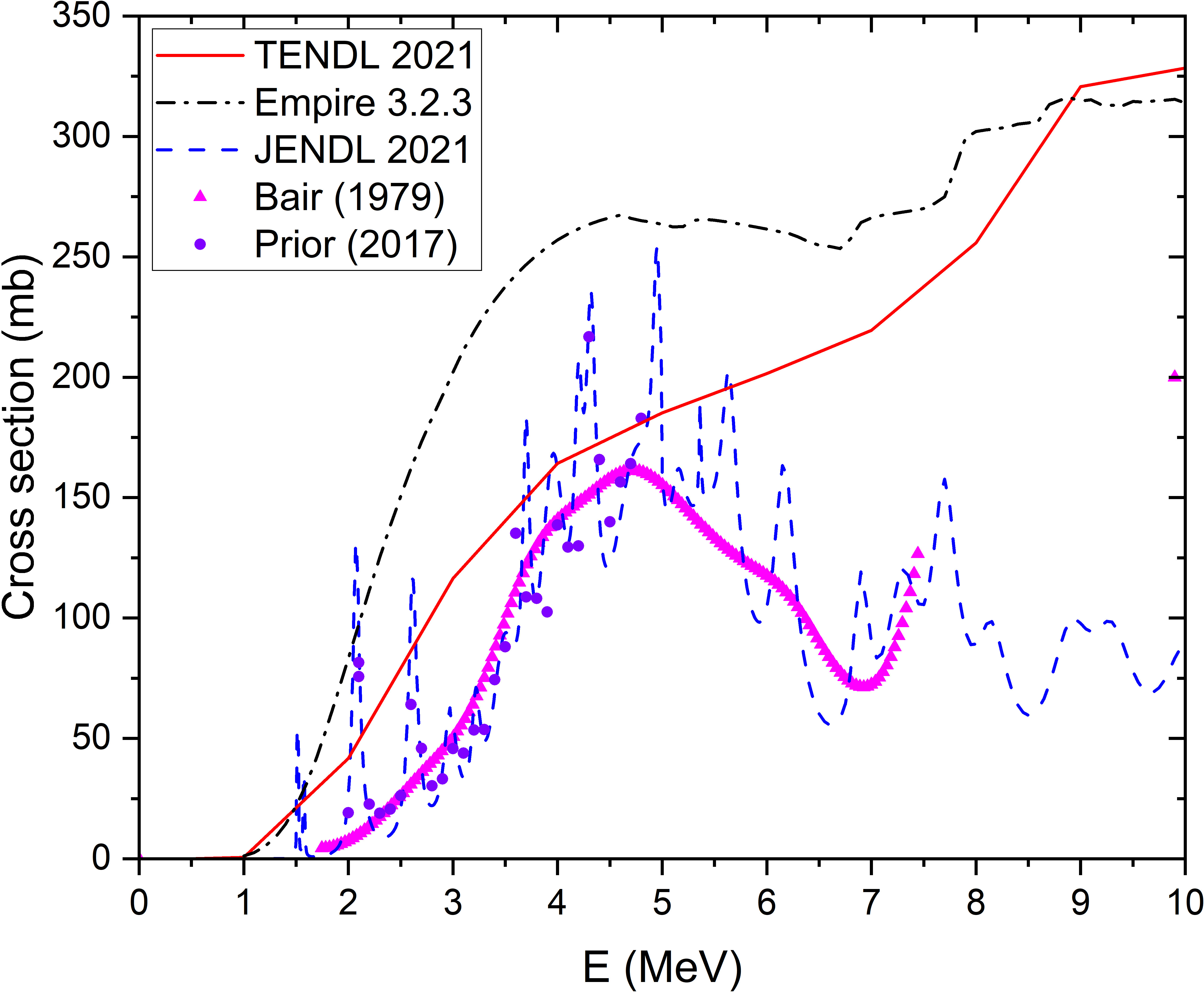}
    \end{minipage}%
    \begin{minipage}{.5\textwidth}
        \includegraphics[width=0.99\linewidth]{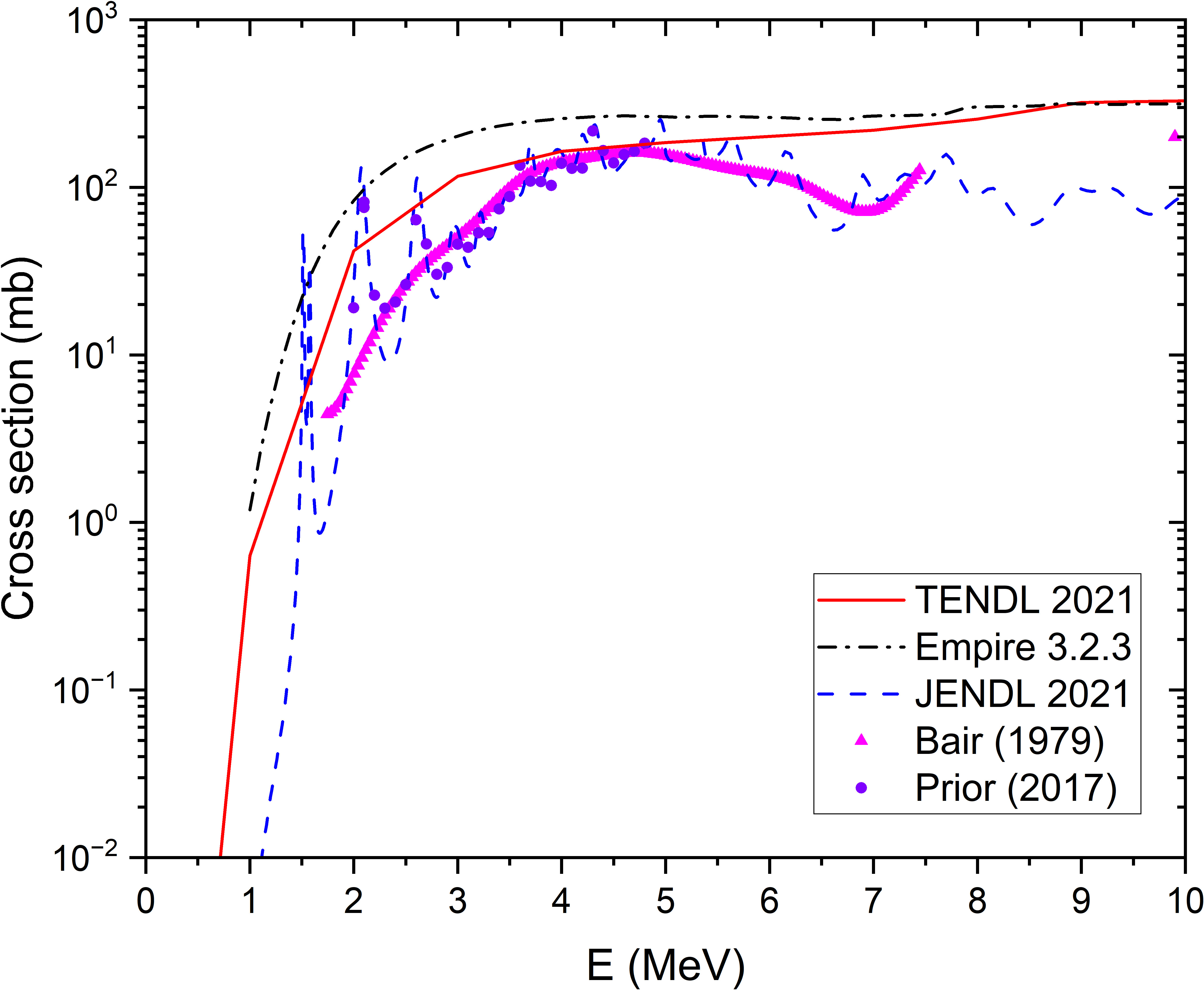}
        \end{minipage}
    \caption{\ce{^{11}B} \alphan \ce{^{14}N} cross-section as a function of energy in linear (left) and logarithmic scale (right) across different orders of magnitude.}
    \label{fig:figB11}
\end{figure}

\begin{figure}[ht]
    \centering
    \begin{minipage}{.5\textwidth} 
        \includegraphics[width=0.99\linewidth]{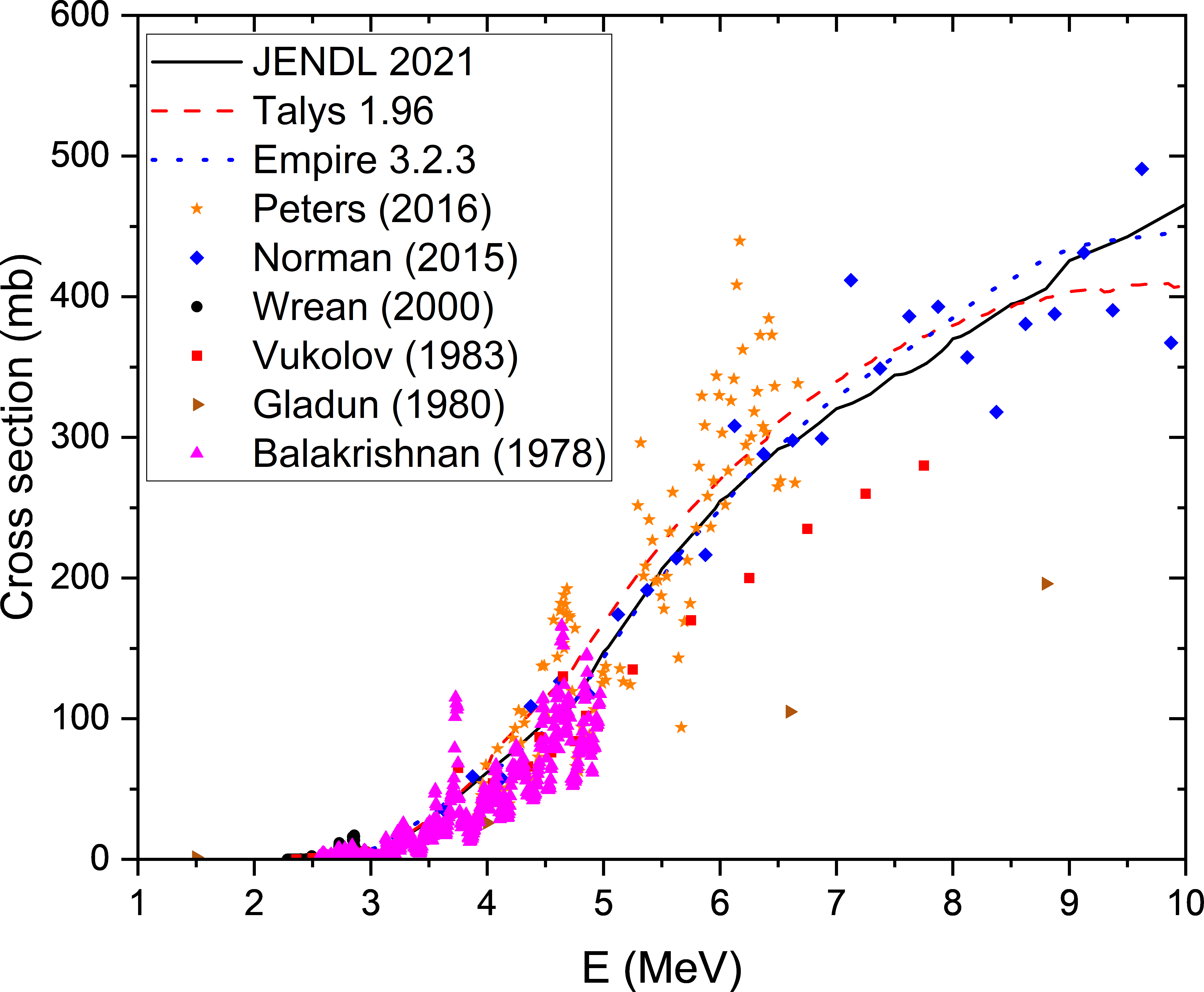}
    \end{minipage}%
    \begin{minipage}{.5\textwidth}
        \includegraphics[width=0.99\linewidth]{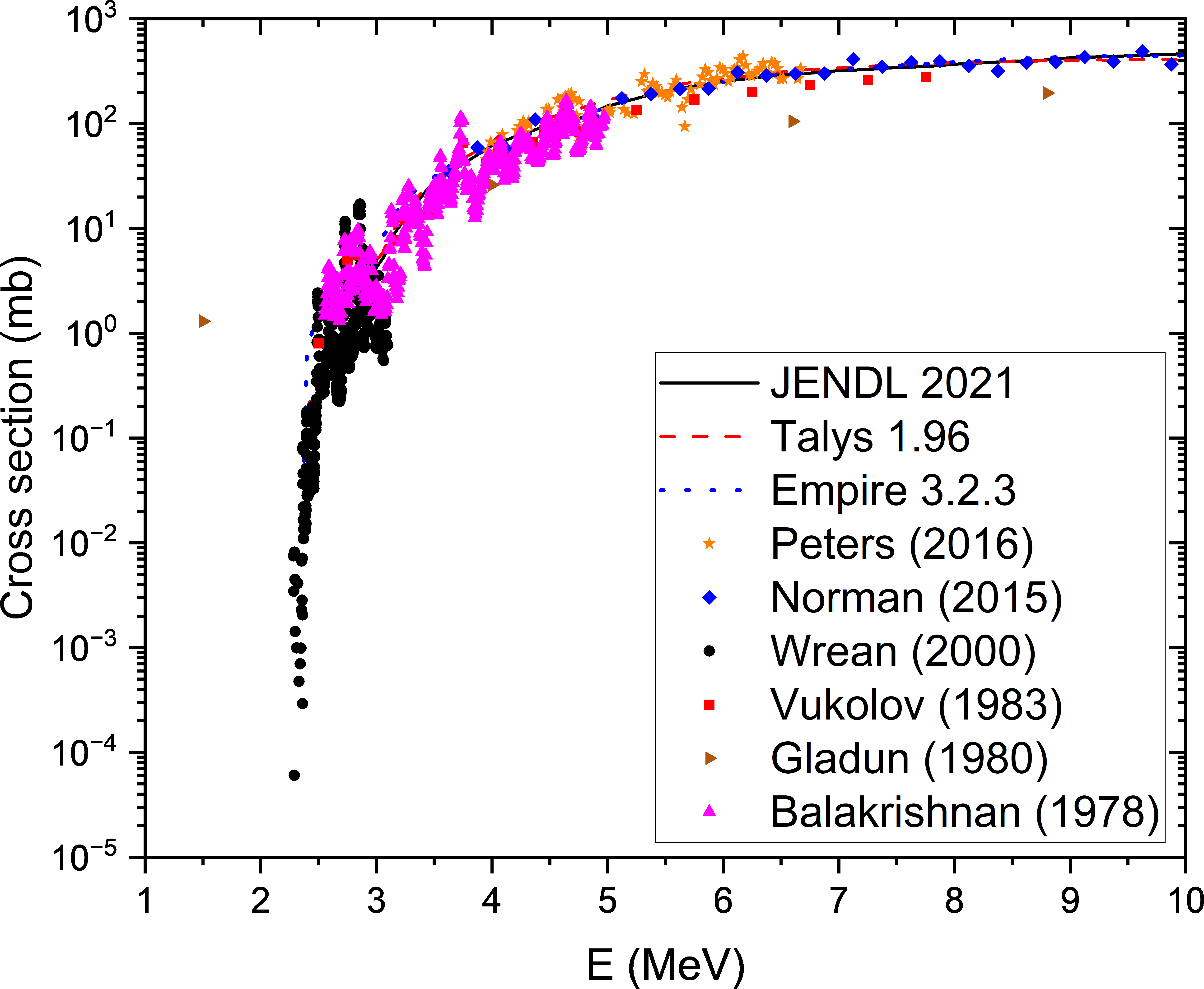}
    \end{minipage}
    \caption{\ce{^{19}F} \alphan \ce{^{22}Na} cross-section as a function of energy in linear (left) and logarithmic scale (right) across different orders of magnitude.}
    \label{fig:figF}
\end{figure}

\begin{figure}[ht]
    \centering
    \begin{minipage}{.5\textwidth}
        \includegraphics[width=0.99\linewidth]{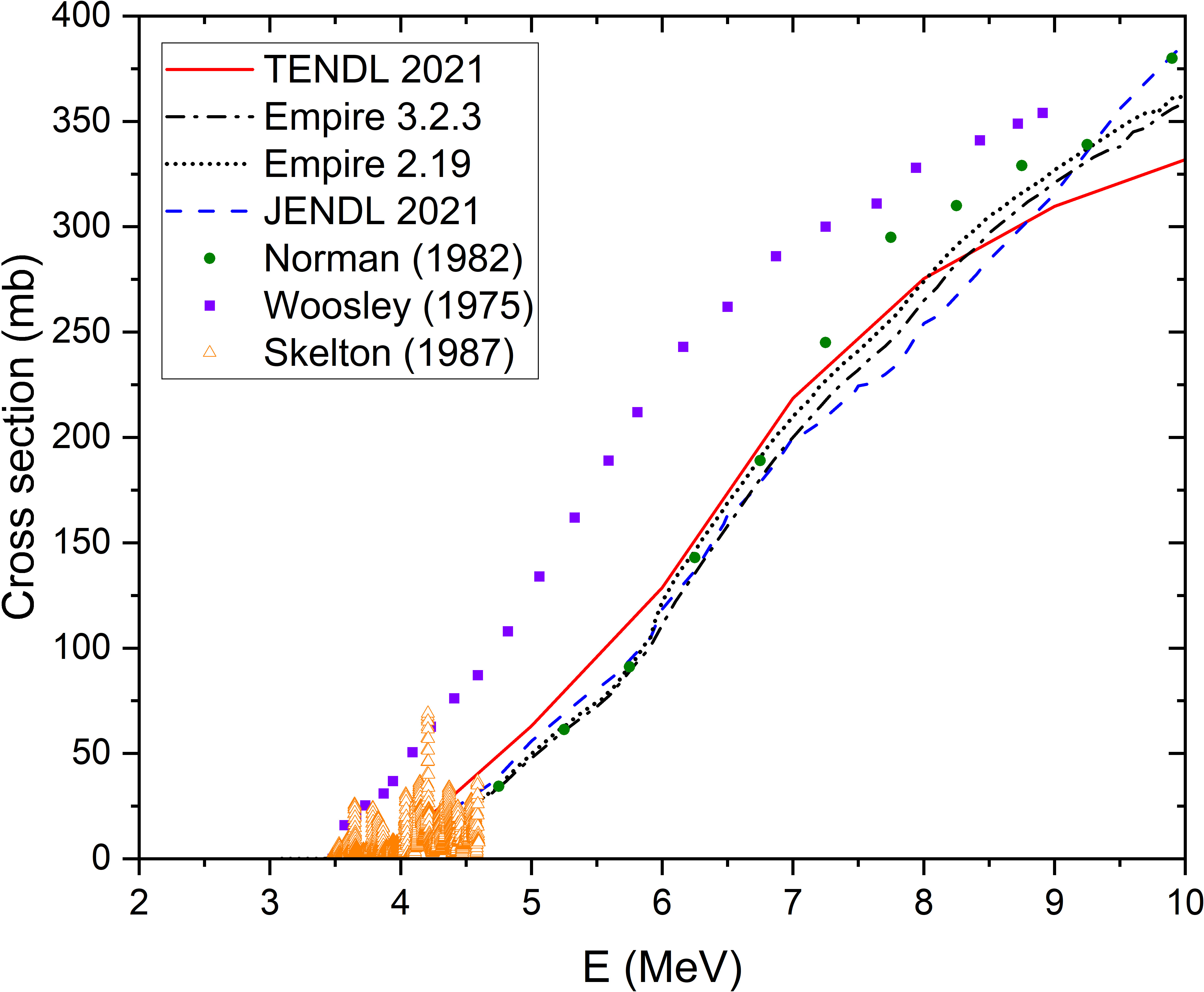}
    \end{minipage}%
    \begin{minipage}{.5\textwidth}
        \includegraphics[width=0.99\linewidth]{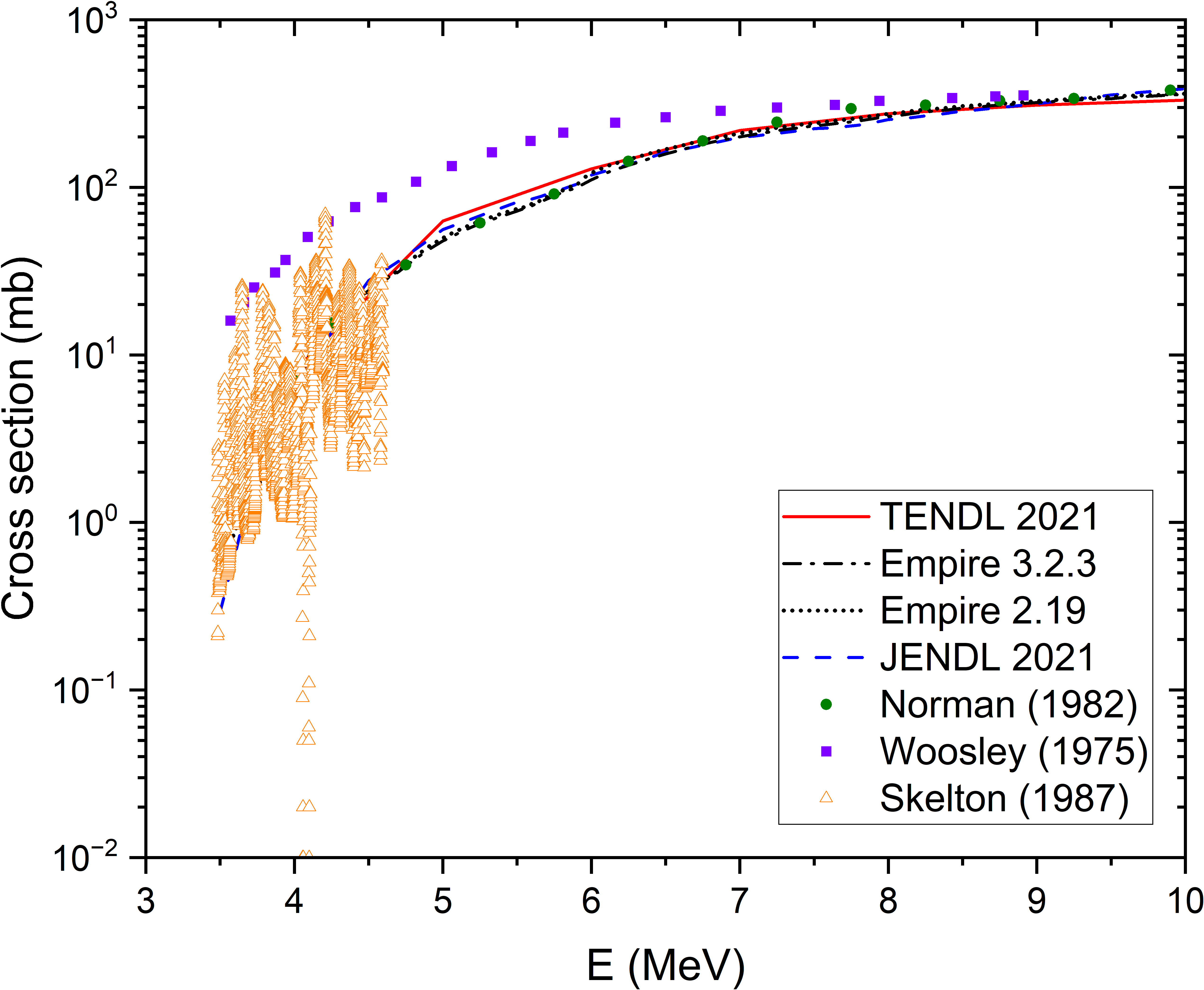}
    \end{minipage}
    \caption{ \ce{^{23}Na} \alphan \ce{^{26}Al} cross-section as a function of energy in linear (left) and logarithmic scale (right) across different orders of magnitude.}
    \label{fig:figN}
\end{figure}

The energy-dependent cross-sections of \alphan\ reactions are nuclide-dependent and can be calculated using nuclear physics codes, such as \EMPIRE\footnote{Currently in version 3.2, available at \url{https://nds.iaea.org/empire}.}~\cite{Herman2007,Herman2015} or \TALYS\footnote{Version 1.96 was used throughout this document. The current version is 2.0, available at \url{https://nds.iaea.org/talys}.}~\cite{koning_modern_2012,Koning2021}, or taken directly from experimental data when available. They can also be taken from available libraries, either evaluated ones like \JENDL~\cite{Iwamoto_jendl-5:2023} or those produced using the toolkits above, such as \TENDL~\cite{koning_tendl_2019}, which is a library of cross-sections obtained with the \TALYS~code. The energy threshold of these reactions is determined by the $Q$-value of the reaction and the Coulomb barrier, which suppresses the reaction probability even if the $\upalpha$-particle energy is above the threshold determined by the $Q$-value. Hence, \alphan\ reactions are important for low- and medium-$Z$ nuclei, while the neutron yield per unit concentration of a radioactive isotope from elements heavier than copper is quite small. 

Although several nuclear reaction databases provide \alphan\ cross-section data, discrepancies among them and gaps in experimental measurements make it difficult to maintain consistency across different computational approaches. Figures \ref{fig:figB10} to \ref{fig:figN} compare cross-section data from different sources for various target nuclei.  
A more structured and accessible repository, integrating both experimental and evaluated data in a standardized format and establishing common reference models and benchmark cases, would significantly improve cross-section availability and usability across various applications.

\subsection{Databases}
\label{sec:Xsec:DB}

The Experimental Nuclear Reaction Data (\EXFOR) library \cite{otuka_towards_2014} (\CSISRS\ in USA) contains an extensive compilation of experimental nuclear reaction data, based on numerical data sets and bibliographical information of \num{22000} experiments since the beginning of nuclear science and it is updated monthly. The associated online database retrieval system provides access to data (by selecting a target, reaction, quantity, or energy range) and different tools for plotting and data comparison. It is publicly available at the websites of the International Atomic Energy Agency Nuclear Data Section\footnote{Available at: \url{https://nds.iaea.org/exfor}.}, the U.S. National Nuclear Data Center\footnote{Available at: \url{https://www.nndc.bnl.gov/exfor}.} and several mirrors. 

The availability of experimental cross-section across the entire energy spectrum of interest can be limited.  Experimental data can be absent for some specific materials.  

On the other hand, only two sets of evaluated \alphan\ cross-section data libraries are available. 

\JENDL\ (Japanese Evaluated Nuclear Data Library) \cite{Iwamoto_jendl-5:2023} provides a library of reaction data for various application fields (the so-called ``general purpose files'') and a set of extra libraries for some particular application field (``special purpose files'')\footnote{Available at: \url{https://wwwndc.jaea.go.jp/jendl/jendl.html}.}. The latest release of the main library was in 2021, named \JENDLv{5}. Among others, it includes information for 795 nuclides for n-induced reactions and 18 nuclides for $\upalpha$-induced reactions in the energy range up to \SI{200}{\MeV} and \SI{15}{\MeV}, respectively.
These 18 nuclides are important mainly in nuclear fuel-cycle applications. They are \ce{^6Li}, \ce{^7Li}, \ce{^9Be}, \ce{^10B}, \ce{^11B}, \ce{^12C}, \ce{^13C}, \ce{^14N}, \ce{^15N}, \ce{^17O}, \ce{^18O}, \ce{^19F}, \ce{^23Na}, \ce{^27Al}, \ce{^28Si}, \ce{^29Si} and \ce{^30Si}. 
The $\upalpha$-particle sub-library includes evaluations of neutron production cross-sections, outgoing neutron energy and angular distribution data utilizing R-matrix theory, statistical model and experimental data depending on the available nuclear structure information for each nuclide \cite{Murata2006}.
\JENDL\ is a truly evaluated library providing recommended nuclear data. Cross-sections in the \JENDL\ library are typically derived from the measurements also included in \EXFOR, with adjustments made to account for disagreements between different measurements and nuclear model codes (mainly using the \mEXIFON\ \cite{etde_545199} and \EGNASH\  \cite{1990ncsc.rept} codes) used to aid the extrapolation of these data.  In some cases, the total neutron production cross-section is adjusted to reproduce measured thick-target neutron yields. All data in version 6 of the Evaluated Nuclear Data File (\ENDFv{6}) format can be retrieved from the dedicated page on the website of the JAEA Nuclear Data Center.

\TENDL\ (\TALYS-based Evaluated Nuclear Data Library) \cite{koning_tendl_2019} is a major nuclear data library that provides the output of the \TALYS\ nuclear model code system for direct use in both basic physics and applications. \TENDL\ contains evaluations for seven types of incident particles (neutrons, protons, deuterons, tritons, \ce{^3He}, $\alpha$-particles, and \grs), for about \num{2800} nuclides and up to energies of \SI{200}{\MeV} for some cases. The 11\textsuperscript{th} version is \TENDLv{2021}, based on both default and adjusted \TALYS\ calculations and data from other sources. First release was in 2008 and the latest ones are \TENDLv{2015}, \TENDLv{2017}, \TENDLv{2019} and \TENDLv{2021}. Since 2015, \TENDL\ is mainly developed at PSI and the IAEA (Nuclear Data Section). Data files are openly available\footnote{{Available under “Libraries” at: \url{https://nds.iaea.org/talys}.}}. In the $\upalpha$ sub-library, tabulated data for total and partial cross-section data and spectra for virtually all nuclides can be found, including the \alphanx\ process; energy is given in \si{\MeV} and cross-section in \si{\milli\barn}. \TALYS\--generated files, formatted according to the \ENDF\ format, are also available. Both total non-elastic cross-section and cross-sections for explicit reaction channels are provided, although some problems are reported for the latter in~\cite{Mendoza:2019vgf} for several nuclides.

\begin{table}[htb]
    \centering
    \caption{Nuclides for which \alphan\ cross-sections are catalogued in the \EXFOR\ and \JENDL\ databases.}
    \begin{tblr}{ccc|ccc|ccc} \toprule
        Nuclide    & EXFOR & JENDL & Nuclide    & EXFOR & JENDL & Nuclide    & EXFOR & JENDL \\ \midrule
        \ce{^6Li}  & \textcolor{Blue}{Yes}   & \textcolor{Blue}{Yes}   & 
        \ce{^25Mg} & \textcolor{Blue}{Yes}   & \textcolor{Red}{No}    & 
        \ce{^59Co} & \textcolor{Blue}{Yes}   & \textcolor{Red}{No}    \\
        \ce{^7Li}  & \textcolor{Blue}{Yes}   & \textcolor{Blue}{Yes}   & 
        \ce{^26Mg} & \textcolor{Blue}{Yes}   & \textcolor{Red}{No}    & 
        \ce{^60Ni} & \textcolor{Blue}{Yes}   & \textcolor{Red}{No}    \\
        \ce{^8Li}  & \textcolor{Blue}{Yes}   & \textcolor{Red}{No}    & 
        \ce{^27Al} & \textcolor{Blue}{Yes}   & \textcolor{Blue}{Yes}   & 
        \ce{^62Ni} & \textcolor{Blue}{Yes}   & \textcolor{Red}{No}    \\
        \ce{^9Be}  & \textcolor{Blue}{Yes}   & \textcolor{Blue}{Yes}   & 
        \ce{^28Si} & \textcolor{Blue}{Yes}   & \textcolor{Blue}{Yes}   & 
        \ce{^63Cu} & \textcolor{Blue}{Yes}   & \textcolor{Red}{No}    \\
        \ce{^10B}  & \textcolor{Blue}{Yes}   & \textcolor{Blue}{Yes}   & 
        \ce{^29Si} & \textcolor{Blue}{Yes}   & \textcolor{Blue}{Yes}   & 
        \ce{^64Zn} & \textcolor{Blue}{Yes}   & \textcolor{Red}{No}    \\
        \ce{^11B}  & \textcolor{Blue}{Yes}   & \textcolor{Blue}{Yes}   & 
        \ce{^30Si} & \textcolor{Blue}{Yes}   & \textcolor{Blue}{Yes}   & 
        \ce{^66Zn} & \textcolor{Blue}{Yes}   & \textcolor{Red}{No}    \\
        \ce{^12C}  & \textcolor{Red}{No}    & \textcolor{Blue}{Yes}   & 
        \ce{^31P}  & \textcolor{Blue}{Yes}   & \textcolor{Red}{No}    & 
        \ce{^68Zn} & \textcolor{Blue}{Yes}   & \textcolor{Red}{No}    \\
        \ce{^13C}  & \textcolor{Blue}{Yes}   & \textcolor{Blue}{Yes}   & 
        \ce{^34S}  & \textcolor{Blue}{Yes}   & \textcolor{Red}{No}    & 
        \ce{^70Zn} & \textcolor{Blue}{Yes}   & \textcolor{Red}{No}    \\
        \ce{^14N}  & \textcolor{Blue}{Yes}   & \textcolor{Blue}{Yes}   & 
        \ce{^35Cl} & \textcolor{Blue}{Yes}   & \textcolor{Red}{No}    & 
        \ce{^72Ge} & \textcolor{Blue}{Yes}   & \textcolor{Red}{No}    \\
        \ce{^15N}  & \textcolor{Blue}{Yes}   & \textcolor{Blue}{Yes}   & 
        \ce{^41K}  & \textcolor{Blue}{Yes}   & \textcolor{Red}{No}    & 
        \ce{^74Ge} & \textcolor{Blue}{Yes}   & \textcolor{Red}{No}    \\
        \ce{^16O}  & \textcolor{Blue}{Yes}   & \textcolor{Red}{No}    & 
        \ce{^45Sc} & \textcolor{Blue}{Yes}   & \textcolor{Red}{No}    & 
        \ce{^76Ge} & \textcolor{Blue}{Yes}   & \textcolor{Red}{No}    \\
        \ce{^17O}  & \textcolor{Blue}{Yes}   & \textcolor{Blue}{Yes}   & 
        \ce{^46Ti} & \textcolor{Blue}{Yes}   & \textcolor{Red}{No}    & 
        \ce{^76Se} & \textcolor{Blue}{Yes}   & \textcolor{Red}{No}    \\
        \ce{^18O}  & \textcolor{Blue}{Yes}   & \textcolor{Blue}{Yes}   & 
        \ce{^48Ca} & \textcolor{Blue}{Yes}   & \textcolor{Red}{No}    & 
        \ce{^86Sr} & \textcolor{Blue}{Yes}   & \textcolor{Red}{No}    \\
        \ce{^19F}  & \textcolor{Blue}{Yes}   & \textcolor{Blue}{Yes}   & 
        \ce{^50Cr} & \textcolor{Blue}{Yes}   & \textcolor{Red}{No}    & 
        \ce{^89Y}  & \textcolor{Blue}{Yes}   & \textcolor{Red}{No}    \\
        \ce{^20Ne} & \textcolor{Blue}{Yes}   & \textcolor{Red}{No}    & 
        \ce{^51V}  & \textcolor{Blue}{Yes}   & \textcolor{Red}{No}    & 
        \ce{^93Nb} & \textcolor{Blue}{Yes}   & \textcolor{Red}{No}    \\
        \ce{^21Ne} & \textcolor{Blue}{Yes}   & \textcolor{Red}{No}    & 
        \ce{^54Fe} & \textcolor{Blue}{Yes}   & \textcolor{Red}{No}    & 
        \ce{^107Ag}& \textcolor{Blue}{Yes}   & \textcolor{Red}{No}    \\
        \ce{^22Ne} & \textcolor{Blue}{Yes}   & \textcolor{Red}{No}    & 
        \ce{^55Mn} & \textcolor{Blue}{Yes}   & \textcolor{Red}{No}    & 
        \ce{^127I} & \textcolor{Blue}{Yes}   & \textcolor{Red}{No}    \\
        \ce{^23Na} & \textcolor{Blue}{Yes}   & \textcolor{Blue}{Yes}   & 
        \ce{^58Ni} & \textcolor{Blue}{Yes}   & \textcolor{Red}{No}    & 
        \ce{^181Ta}& \textcolor{Blue}{Yes}   & \textcolor{Red}{No}    \\ \bottomrule
    \end{tblr}
    \label{tab:xsdata}
\end{table}

\begin{table}[htb]
    \centering
    \caption{Mid-$Z$ range naturally occurring isotopes for which there are currently no measured \alphan\ cross-sections in the \SIrange{4}{10}{\MeV} energy range cataloged in \EXFOR. Elements for which * is listed for the mass number have no data available for any of their naturally existing isotopes. Red entries (all except \ce{Ni} and \ce{V}) are those for which missing isotopes collectively contribute at least \SI{10}{\percent} of the element’s total neutron yield. }
    \begin{tblr}{rrrrrrr}\toprule
        {\color{red} \ce{^{32,33,36}S}} & {\color{red} \ce{^37Cl}} & {\color{red} \ce{^*Ar}} & {\color{red} \ce{^{39,40}K}} & {\color{red} \ce{^{42-44,46}Ca}} & {\color{red} \ce{^{47,49,50}Ti}} & \ce{^50V} \\
        {\color{red} \ce{^{52-54}Cr}} & {\color{red} \ce{^{56-58}Fe}} & \ce{^{61}Ni} & {\color{red} \ce{^67Zn}} & {\color{red} \ce{^73Ge}} & {\color{red} \ce{^{74,77,78,80,82}Se}} & {\color{red} \ce{^{*}Br}} \\
        {\color{red} \ce{^*Kr}} & {\color{red} \ce{^*Rb}} & {\color{red} \ce{^{84,87,88}Sr}} & {\color{red} \ce{^*Zr}} & {\color{red} \ce{^{95,96-98}Mo}} & {\color{red} \ce{^{96,99-102,104}Ru}} & {\color{red} \ce{^*Rh}} \\
        {\color{red} \ce{^*Pd}} & {\color{red} \ce{^*Cd}} & {\color{red} \ce{^113In}} & {\color{red} \ce{^*Sn}} & {\color{red} \ce{^{120,122-126,128}Te}} & \\
        \bottomrule
    \end{tblr}
    \label{tab:missing_xsdata}
\end{table}

Nuclides for which experimentally measured or evaluated cross-sections exist and overlap with the radiogenic $\upalpha$s energy range are summarized in Table~\ref{tab:xsdata}. While the available data cover most naturally occurring nuclides, there are several elements that lack \alphan\ cross-section measurements and are relevant for low-background experiments. These elements, commonly used in detector structural materials or as targets, are typically in the mid-$Z$ range and include metals and noble gases.

Nuclides lacking cross-section data with atomic numbers up to that of iodine are summarized in Table~\ref{tab:missing_xsdata}. To quantify the impact of missing data, the neutron yield of each element’s naturally occurring isotopes has been computed using \NeuCBOT\ \cite{Gromov_2023} (see Section~\ref{sec:Tools}) for \SI{10}{\MeV} $\upalpha$-particles, employing cross-sections from \TALYS. If cross-section measurements are missing for isotopes that collectively contribute at least \SI{10}{\percent} of the total neutron yield, the element is highlighted in red in Table~\ref{tab:missing_xsdata}.

Among the elements analyzed, Ar, Br, Kr, Rb, Rh, Pd, Cd, and Sn lack cross-section data for all of their naturally occurring isotopes.  

For elements with atomic numbers greater than that of iodine, no cross-section measurements are available, with the sole exception of \ce{^181Ta}. However, the total neutron yield from such heavy nuclides is typically much lower than that of lighter elements (e.g. \NeuCBOT\ predicts an \alphan\ yield of \num{7e-11} neutrons per $\upalpha$-particle for natural xenon at \SI{10}{\MeV}).  

A detailed comparison of the results from \JENDLan\ and \TENDLv{2014}, \TENDLv{2015}, and \TENDLv{2017} is provided in~\cite{Mendoza:2019vgf}; \TENDL\ is found to yield larger neutron production cross-sections in most cases, with a few exceptions.

\subsection{Models}
\label{sec:Xsec:Models}

Depending on the kinetic energy $E_\upalpha$ of the incident $\upalpha$ particle, various reactions can occur\footnote{See e.g.\ \cite{Bertulani2019} for an introduction into nuclear reaction theory.}: below the Coulomb barrier of the target nucleus $X$, the incident particle may scatter elastically $X(\upalpha, \upalpha)X$, at higher energies, inelastic scattering $X(\upalpha, \upalpha')X^*$ may leave the target nucleus $X$ in an excited level. At low energies and especially for deformed nuclei this may be a collective excitation e.g.\ rotation or vibration. At higher energies, stripping reactions may cause a neutron emission. Above the Coulomb barrier, the incident $\upalpha$-particle can be captured: after a pre-equilibrium phase, the incident nucleons are thermalized, resulting in the creation of a compound nucleus ${}^{A+2}_{Z+2}X^*$ excited by the added energy: both the kinetic energy of the $\upalpha$ and the binding energy $S_\upalpha$. The compound nucleus may de-excite via several decay channels: e.g.\ either via $\upgamma$ transition to the ground state or, if the incident energy $E_\upalpha+S_\upalpha$ exceeds a multiple of the neutron separation energy $S_\mathrm{n}$, via the emission of one or more neutrons together with one or more \grs, i.e. \alphan, ($\upalpha, \mathrm{n} \upgamma$), ($\upalpha, 2\mathrm{n} \upgamma$), ($\upalpha, 2\mathrm{n} 2\upgamma$), \ldots, ($\upalpha, x \mathrm{n} y \upgamma$) with the related exclusive excitation functions  (probabilities of transitions to excited states) $\sigma_{\upalpha,x \mathrm{n} y \upgamma}(E_\upalpha)$. Of interest for us is the excitation function for at least one neutron in the decay channel: $\sigma_\mathrm{\upalpha,n}(E_\upalpha)=\sum_x \sum_y \sigma_{\upalpha,x \mathrm{n} y \upgamma}(E_\upalpha)$.

In the range of resolved resonances on individual levels, i.e.\ where the width of the peaks in $\sigma_\mathrm{\upalpha,n}(E_\upalpha)$ is smaller than their spacing, the reaction has to be described by R-matrix theory (see e.g.\ \cite{Descouvemont2010}). Once the spacing drops below the peak's width in the range of unresolved resonances, the reaction can be treated in a statistical way. As the spacing depends on the level density that increases with the mass of the nucleus, the energy below which the statistical treatment breaks down increases with decreasing nuclide mass.

The formation and decay of the compound nucleus can be statistically treated by the  Hauser-Feshbach theory \cite{Hauser1952}. Generally, the exit channels are treated as independent of the entrance channel; remaining correlations between them at low energy are treated by width fluctuation corrections. The transmission coefficients for the entrance and exit channels are commonly calculated within an optical model (see e.g.\ \cite{Carlson2001,Capote2009}), i.e.\ describing the elastic scattering and the inelastic scattering via the real and imaginary parts of a complex optical model potential (OMP), respectively. The basic OMPs for spherical nuclides may be generalized for deformed nuclides to coupled-channels (CC) optical models. For weak coupling, the distorted wave-Born approximation (DWBA) can be applied to the CC OMP. Albeit by definition the statistical treatment cannot describe the resolved resonances, it can describe the average behaviour of the relevant physical quantities in this energy range \cite{Rochman2013}: based on these average values and the underlying distributions, it is possible to randomly sample resonances to populate the range of resolved resonances. Albeit the individual resonances created this way depend on the used random seeds, their average behaviour is physically sound.

Concerning the OMPs, they can be broadly divided in two classes \cite{Capote2009}: phenomenological or microscopic OMPs. For both classes, models may include dispersion, i.e.\ do not treat the real and imaginary parts of the complex potential as independent, and hence reducing ambiguities in their parameters. The microscopic OMPs rely on folding the target matter distribution with OMP in nuclear matter, whereas phenomenological OMPs parameterize experimental data, either for an individual nuclide (\emph{local} OMP) or for a larger set of nuclides (\emph{global} OMP). Global phenomenological OMPs are optimized in such a way that, on average, they reproduce cross-sections close to the experimental ones for all non-actinide nuclides. This means there are some nuclides, for example, \ce{^13C}, for which the calculated numbers are not very reliable, and extra tuning or even changing to a dedicated local OMP is desirable. For a discussion of suitable OMPs, see section~\ref{sec:Uncertainty:xsec}.

Nuclear reaction codes that implement the relevant models, and which are widely used and actively maintained, are \TALYS\ and \EMPIRE. Both use external tools for OMP calculations: \TALYS\ relies on ECIS-06\footnote{Available at: \url{https://nds.iaea.org/RIPL-3/codes/ECIS}.} \cite{Raynal1972}; also \EMPIRE\ can use ECIS-06 but offer alternatively OPTMAN in the unpublished version 12\footnote{The older version 10 is available at: \url{https://nds.iaea.org/RIPL-3/codes/OPTMAN}.} \cite{Soukhovitskil2008} which has an improved treatment of rotational levels for CC calculations. As both \TALYS\ and \EMPIRE\ are based on similar statistical models, their results are not significantly different from each other. A comprehensive comparison between the two codes, \TALYS\ and \EMPIRE, and experimental data for several key nuclides has been reported in Ref.~\cite{zakhary,kudryavtsev_neutron_2020}.

%% file: TexFiles/Tools.tex
\section{Neutron yield calculation tools}
\label{sec:Tools}

The codes to calculate neutron yields and spectra require a number of inputs, such as the energy-dependent cross-sections of \alphan\ reactions, transition probabilities to excited states (i.e. ``excitation functions'') and stopping power of $\upalpha$-particles in different materials. These codes also need material composition where a particular reaction occurs, $\upalpha$-energy or concentration of radioactive nuclei in the material.

The excitation functions are typically derived from calculations similar to those used for cross-sections. Measurements are scarce and usually contain only data from the first few excited states.

The $\upalpha$-particles produced in radioactive decay quickly lose energy in a material through ionization and excitation of atoms, thereby reducing the probability of neutron production. While energy loss of $\upalpha$-particles is assumed to be well understood and is taken into account in all codes dealing with neutron yield calculation, uncertainties in models can sometimes be important. In particular, the breakdown of Bragg's additivity law in composite low-$Z$ materials may result in uncertainties up to \SI{50}{\percent}~\cite{thwaites_braggs_1983}.

\subsection{Stopping power calculation}
\label{sec:Tools:SP}

The naturally occurring radioactive decay of some actinides can produce $\upalpha$-particles. \ce{^238U}, \ce{^232}Th and \ce{^235U} are present in the Earth's crust at the $\approx$ppm (particle per million) level and have decay chains in which several $\upalpha$-decays are produced in series. Their initial energy is between \SIrange{4}{10}{\MeV}, so the range of energies \SIrange{0}{10}{\MeV} needs to be considered to account for the energy loss in the material before the interaction.
Nowadays, there exist several programs for the calculation of the stopping power in various materials, such as \ASTAR\ and \SRIM. Additionally, all published experimental data on stopping powers are compiled and disseminated through the IAEA Stopping Power Database\footnote{Available at: \url{https://nds.iaea.org/stopping}.}.

The NIST code \ASTAR\ \cite{astar} calculates the stopping power of $\upalpha$-particles in 74 media based on the methods described in the ICRU~49 report~\cite{ICRU49}. The stopping power is given in the energy range from \SI{1}{\keV} to \SI{1}{\GeV}. Tables are differentiated for solid and gaseous targets as differences up to \SI{10}{\percent} are expected~\cite{thwaites_braggs_1983}. In the high energy range ($E_\upalpha>\SI{2}{\MeV}$) the stopping power is calculated using Bethe's formula and includes the shell, Barkas and Bloch, and the density-effect corrections, while, in the low energy range, models based on experimental data are used. When element-specific measurements are lacking values are interpolated among the existing ones~\cite{astar}. For composite material without experimental data, the cumulative stopping power is calculated by linearly adding the stopping power of each element in the compound (Bragg's additivity law).

The Stopping and Range of Ions in Matter (\SRIM) code~\cite{srim}, on the other hand, uses a quantum mechanical treatment of projectile-target atoms collisions to calculate the stopping power and range of various ions (projectiles), including $\upalpha$-particles, in matter (target atoms). In the case of compound materials, the code takes into account the energy loss due to the bonding electrons, correcting Bragg's additivity law. This is shown to be particularly important for low-$Z$ materials, such as hydrocarbons, where chemical bonding effects can result in up to \SI{50}{\percent} deviations between data and calculations \cite{thwaites_braggs_1983}.
\SRIM\ has the flexibility to manually input the target's composition based on its molecular weights in addition to the existing target material's list. Similar to the \ASTAR\ database, \SRIM\ allows the selection of a solid or gaseous target. For $\upalpha$-particles the range is extended up to \SI{8}{\GeV}.

Several authors have performed comparisons between the \ASTAR\ and \SRIM\ calculations for a range of materials~\cite{kumar2018, jassim2017, PIGNI2016147, HANSSON2020100734}. The two calculations are generally in good agreement, typically within a few percent. When compared to data, the result is highly dependent on the material. The \ASTAR\ program reports an uncertainty of approximately \SIrange{1}{2}{\percent} for single elements and \SIrange{1}{4}{\percent} for compound materials within the energy range relevant for radiogenic $\upalpha$-particles (\SIrange{2}{8}{\MeV}). For lower energies ($E<\SI{1}{\MeV}$), the uncertainty increases as energy decreases, reaching about \SI{30}{\percent} for \SI{1}{\keV} $\upalpha$-particles~\cite{astar}. The authors of \SRIM\ report that \SI{89}{\percent} of the calculations performed with the code agree within \SI{10}{\percent} with the experimental data for $\upalpha$-particles~\cite{srim}.

The tools presented in this paper for the neutron yield calculation either use \SRIM\ outputs (\NeuCBOT), or implement the stopping power coefficients tabulated in Refs.~\cite{ziegler1977, perry1981} (\SOURCESF). \SaGFN\ uses \GeantF\ stopping power tables which, for $\upalpha$ energies below \SI{8}{\MeV}, are based on the ICRU report and \ASTAR.

\subsection{SOURCES4}
\label{sec:Tools:S4}

The code \SOURCESF~\cite{sources4a,wilson_sources_2005} uses the libraries of $\upalpha$-emission lines from radioactive nuclides, cross-sections of \alphan~reactions either from calculations or experimental data, excitation functions and energy losses of $\upalpha$-particles in different materials, to calculate the neutron production rate and energy spectra of emitted neutrons. The most recent version is \SOURCESFC~\cite{wilson_sources_2005} but for historical reasons, the older version \SOURCESFA~\cite{sources4a} is used by some collaborations. The release notes from the authors and previous tests showed that, if the same cross-sections and branching ratios are used in both versions of the code, there is no difference between the results for almost all nuclides. 

The code was modified to extend the $\upalpha$-energy range from \SI{6.5}{\MeV} (as in the original version) to about \SI{10}{\MeV}~(see \cite{tomasello_calculation_2008} for the modifications of the \SOURCESFA~code). More cross-sections and branching ratios calculated using the \EMPIREv{2.19}~\cite{empire}, \EMPIREv{3.2.3} and \TALYS~\cite{koning_modern_2012} codes were added to the  library of the \SOURCESF\ code covering the range of $\upalpha$-energies up to \SI{10}{\MeV}~\cite{carson_neutron_2004,tomasello_calculation_2008,lemrani,tomasello2010,tomasello-thesis,kudryavtsev_neutron_2020}. A comparison of cross-sections from \EMPIREv{2.19} with experimental data was published in Refs.~\cite{tomasello_calculation_2008,tomasello-thesis} and the results of neutron yield calculations with modified \SOURCESFA~were used for a number of dark matter experiments (see, for example, Refs.~\cite{eureca,edelweiss,lz,xenon1t}). The accuracy of the calculation was estimated to be about \SI{20}{\percent} based on the comparison of neutron yields obtained with different sets of cross-sections~\cite{tomasello_calculation_2008}. 

The user input to \SOURCESFA~includes material composition (where an $\upalpha$-source is located), isotopic composition for each element (only isotopes with cross-sections present in the code library can be included) and either the energy of the $\upalpha$-particle or the radioactive nuclide (or several nuclides in the case of decay chains, for instance) with the number of atoms in a sample. For application in low-background experiments an option of the thick target neutron yield is used, meaning that the size of the sample is much bigger than the range of $\upalpha$s and edge effects can be neglected. 

The output of \SOURCESFA~includes several files that return the neutron yield and spectra for the sum of the ground state and all excited states, as well as neutron spectra for individual states. The neutron yield is also calculated for each $\upalpha$-emitter in a decay chain and for every type of nuclide present in the material sample. \SOURCESFA/\SOURCESFC~do not calculate \gr~production, but the total energy transferred to \grs~can be calculated from the energy of the excited states.

Cross-sections from recent versions of the nuclear physics codes 
\TALYS~\cite{koning_modern_2012} and \EMPIREv{3.2.3} \cite{empire} have become available over the past few years and were added to the \SOURCESFA~libraries. The comparison between early neutron yield calculations with cross-sections from \EMPIREv{2.19}, \TALYSv{1.9} and \EMPIREv{3.2.3} has recently been published~\cite{kudryavtsev_neutron_2020}. 

\SOURCESFA~libraries are regularly updated as described in Ref.~\cite{kudryavtsev_neutron_2020}. Recent development includes optimisation of the cross-sections and branching ratios used in the calculations by selecting a combination of the experimental data and the most reliable model for a particular nuclide~\cite{vk2022,parvu2024}. For low-$Z$ materials, measurements are used where available since none of the codes based on statistical models can reliably predict \alphan~cross-sections. These data are complemented by the calculations from \EMPIRE~or \TALYSv{1.9} to extend the cross-sections to higher energies and to obtain branching ratios usually unavailable from limited data sets. The optimisation is being validated by comparing the neutron yield as a function of $\upalpha$ energy with the available data.

\subsection{NeuCBOT}
\label{sec:Tools:NCBT}

\NeuCBOT\footnote{Available at: \url{https://github.com/shawest/neucbot}.} (the Neutron Calculator Based On \TALYS) allows the user to specify a material composition --- either by element, assuming natural abundances, or by nuclide --- and a material contamination level \cite{Westerdale:2017kml}. Material contamination can be described either with a list of $\upalpha$-particle energies or $\upalpha$-emitting nuclides, all weighted by their desired relative abundance. 
\NeuCBOT\ then simulates $\upalpha$-particles slowing down in the material and integrates over the \alphan\ cross-section and emitted neutron spectrum at each step.

Stopping powers are read from a library generated by \SRIM\ and summed together assuming Bragg's additivity law. If the user specified contamination levels by the $\upalpha$-emitting nuclides, a local database is built by retrieving \ENSDF\ files from the NNDC NuDat database, and $\upalpha$ energies and branching ratios are retrieved.
The \alphan\ cross-sections and emitted neutron spectra are drawn from a \TALYS-generated database. This database can be generated locally if the user has a local version of \TALYS\ installed, or it can be retrieved from a remote database, which is generated for all naturally occurring nuclides for $\upalpha$ energies up to \SI{10}{\MeV}. To reduce disk space used in the latter case, the databases are only retrieved for each needed element and stored locally. \NeuCBOTv{1} uses a database generated with \TALYSv{1.6}, while \NeuCBOTv{2} \cite{Gromov:2023iuh} uses \TALYSv{1.95}. 
Generally, \NeuCBOTv{1} \alphan\ yields were found to agree with similar calculations uses \JENDL\ to within about \SI{30}{\percent} in most cases, with a few cases where bigger disagreement is seen, and the yields calculated by \NeuCBOT\ are typically higher. For \NeuCBOTv{2}, a significant decrease in yield is found, bringing the values into closer agreement with numerical integrals over \JENDL\ data.
A more recent version, \NeuCBOTv{3} allows the user to select evaluated cross-sections from the \JENDL\ library where available, rather than those simulated by \TALYS.
Additional upgrades are currently under development, including calculations of correlated \gr\ yields and correlations between outgoing neutron energy and energy lost by the $\upalpha$-particle prior to capture.


\subsection{SaG4n}
\label{sec:Tools:SGN}

Until recently, general-purpose radiation transport simulation codes like \GeantF~\cite{GEANT4:2002zbu} were not able to calculate \alphan\ yields with sufficient accuracy due to the difficulty of realistically modeling low-energy $\upalpha$-reactions. Since version 10.2, released in 2015, \GeantF\ has incorporated the so-called ParticleHP module \cite{Allison:2016lfl}, which uses data libraries originally written in \ENDFv{6} format to handle non-elastic nuclear reactions of low-energy (\SI{<200}{\MeV}) charged particles. These data libraries provide detailed information on nuclear interactions, including reaction cross-sections and secondary particle production. This advancement allows for modeling \alphan\ reactions with much higher precision than was previously possible using theoretical models implemented directly in the code.

\SaGFN~\cite{Mendoza:2019vgf,Pesudo:2020bgh}, a \GeantF-based code, was specifically developed for the calculation of \alphan\ neutron yields. This code\footnote{Available at: \url{https://github.com/UIN-CIEMAT/SaG4n}.} is compatible with Geant4.10.6 and later versions. \SaGFN\ employs explicit transport of incident $\upalpha$-particles through the material using stopping power tables from \GeantF\ (G4EmStandardPhysics-option4), which for $\upalpha$ energies below 8 MeV are based on data tables from ICRU and \ASTAR. Instead of relying solely on neutron production cross-sections, neutrons are produced individually as nuclear reactions occur. This approach enables the simulation of both neutron production and transport within \GeantF, providing detailed information on each individual \alphan\ reaction. This level of detail is essential for accurately calculating background events caused by \alphan\ neutrons, particularly in rare event search experiments. While \SaGFN\ requires more computational time than other codes, this is mitigated by well-known biasing techniques that enhance the neutron production rate and reduce the overall computational cost.

\SaGFN\ provides flexibility in selecting data libraries and allows users to modify input and output files as needed. The primary data library is based on \JENDLan, with \TENDL\ serving as a fallback for nuclides not included in \JENDLan. The code output includes the initial position and momentum of the incident $\upalpha$-particle, the position and momentum of the produced neutron (and $\upgamma$-rays, if generated), and the ``weight'' of the event. This weight compensates for the biasing techniques applied to enhance the neutron production rate for \alphan\ reactions.

In addition to the detailed simulation of particle generation, \SaGFN\ allows to simulate highly complex material geometries, far more advanced than those supported by other codes. Being based on \GeantF, it benefits from a wide range of tools for defining, verifying, and importing geometries. These tools include robust capabilities for visualizing the geometry, checking overlaps or inconsistencies, and importing CAD-based models or predefined detector components, ensuring accurate and efficient calculations.

This feature significantly improves the accuracy of results, particularly in scenarios involving interfaces between different materials. For instance, this is critical for modeling surface $\upalpha$s from, for example, radon progeny plate-out, which strongly influence background calculations and systematic uncertainties in low-background experiments.

Moreover, \SaGFN\ can, in principle, generate neutrons and $\upgamma$-rays in coincidence, along with other secondary particles emitted in the same nuclear reaction. However, the reliability of the results strongly depends on the specific nucleus where the nuclear reaction occurs. This is because, for the results to be reliable, the information in the data libraries must be encoded in a specific way that preserves the correlations between the secondary particles.

\begin{table}
\caption{Neutron yield from \alphan~reactions in the \ce{^238U} and \ce{^232Th} decay chains in several materials as calculated by different codes. (\SI{1.07}{\percent} of \ce{^13C} has been assumed for natural carbon.)}
\label{table:code-comparison}
\centering
\begin{tblr}{
	colspec={
	l
	Q[si={table-format=1.2e-2,table-number-alignment=center},c]
	Q[si={table-format=1.2e-2,table-number-alignment=center},c]
	Q[si={table-format=1.2e-2,table-number-alignment=center},c]
	},
	row{1}={guard},
	row{2}={guard}
}
\toprule
\SetCell[r=2]{h} Material & \SetCell[c=3]{c} Neutron yield, g$^{-1}\,$s$^{-1}\,$ppb$^{-1}$, as calculated by & &\\\cmidrule{2-4}
 & SOURCES4 & NeuCBOT v-3.0 (JENDL) & SaG4n (JENDL) \\\midrule
Carbon, \ce{^238U} & 1.35E-11 & 1.25e-11 & 1.35e-11 \\
Carbon, \ce{^232Th} & 5.53E-12 & 5.21e-12 & 5.65e-12 \\
Fluorine, \ce{^238U} & 1.34e-9 & 1.00e-9 & 1.21e-9 \\
Fluorine, \ce{^232Th} & 5.35e-10 & 4.45e-10 & 5.23e-10 \\
Aluminium, \ce{^238U} & 1.58e-10 & 1.41e-10 & 1.52e-10 \\
Aluminium, \ce{^232Th} & 8.26e-11 & 7.55e-11 & 8.11e-11 \\
\bottomrule
\end{tblr}
\end{table}

\subsection{Comparison between the codes}
\label{code-comparison}

\begin{figure}[ht!]
    \centering
    \includegraphics[width=\textwidth]{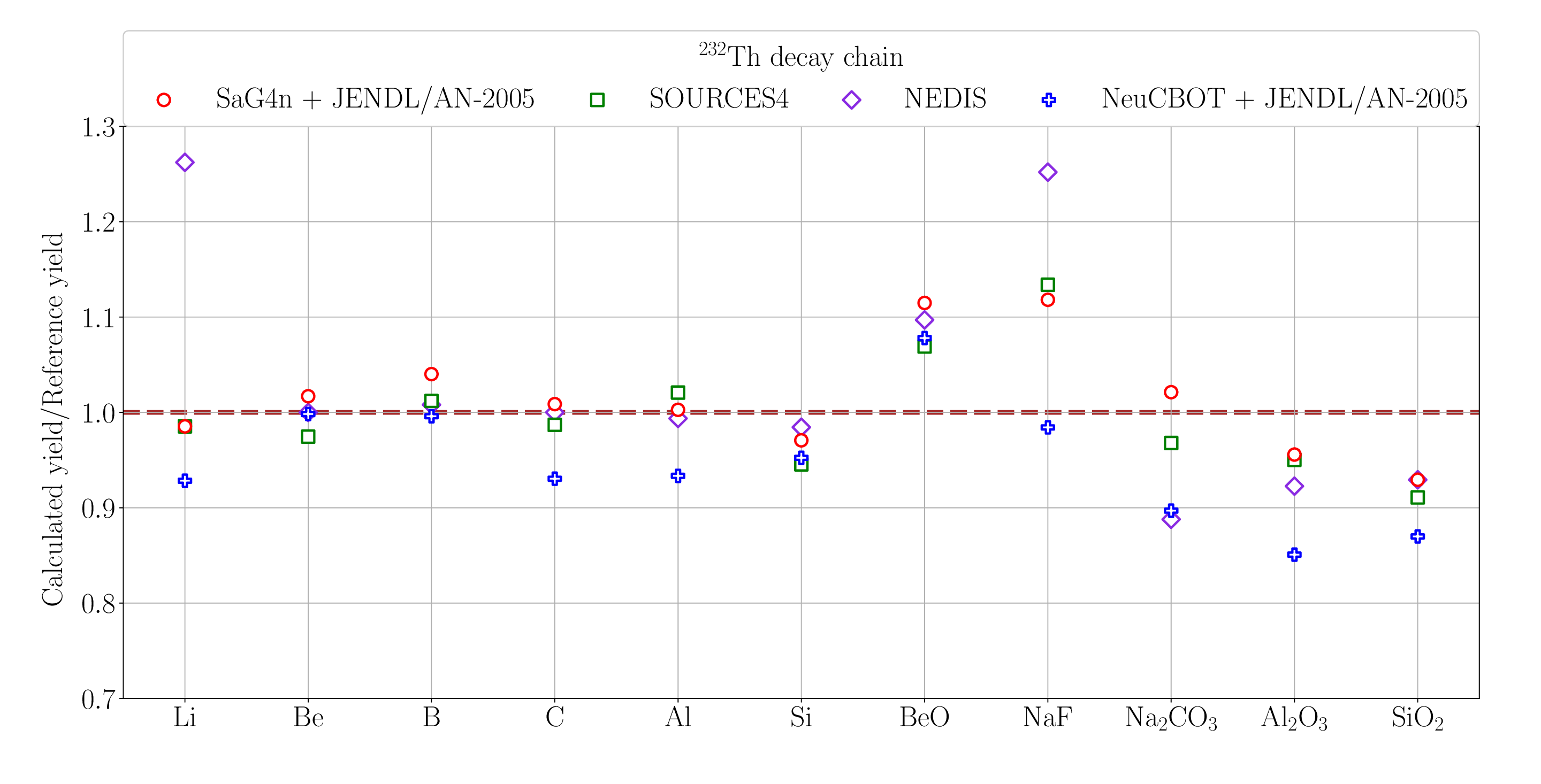}
    \caption{Comparison of neutron yields calculated for light nuclei by applying different codes, including the upgraded version of \NeuCBOT. The numerical values are normalized to evaluated data from various $\upalpha$-beam measurements \cite{fernandes} in order to demonstrate agreement between the calculations and measurements. The comparison in question is performed only for the thorium series.}
    \label{fig:comp_all_codes_232th}
\end{figure}

\begin{figure}[ht]
    \centering
    \includegraphics[width=\textwidth]{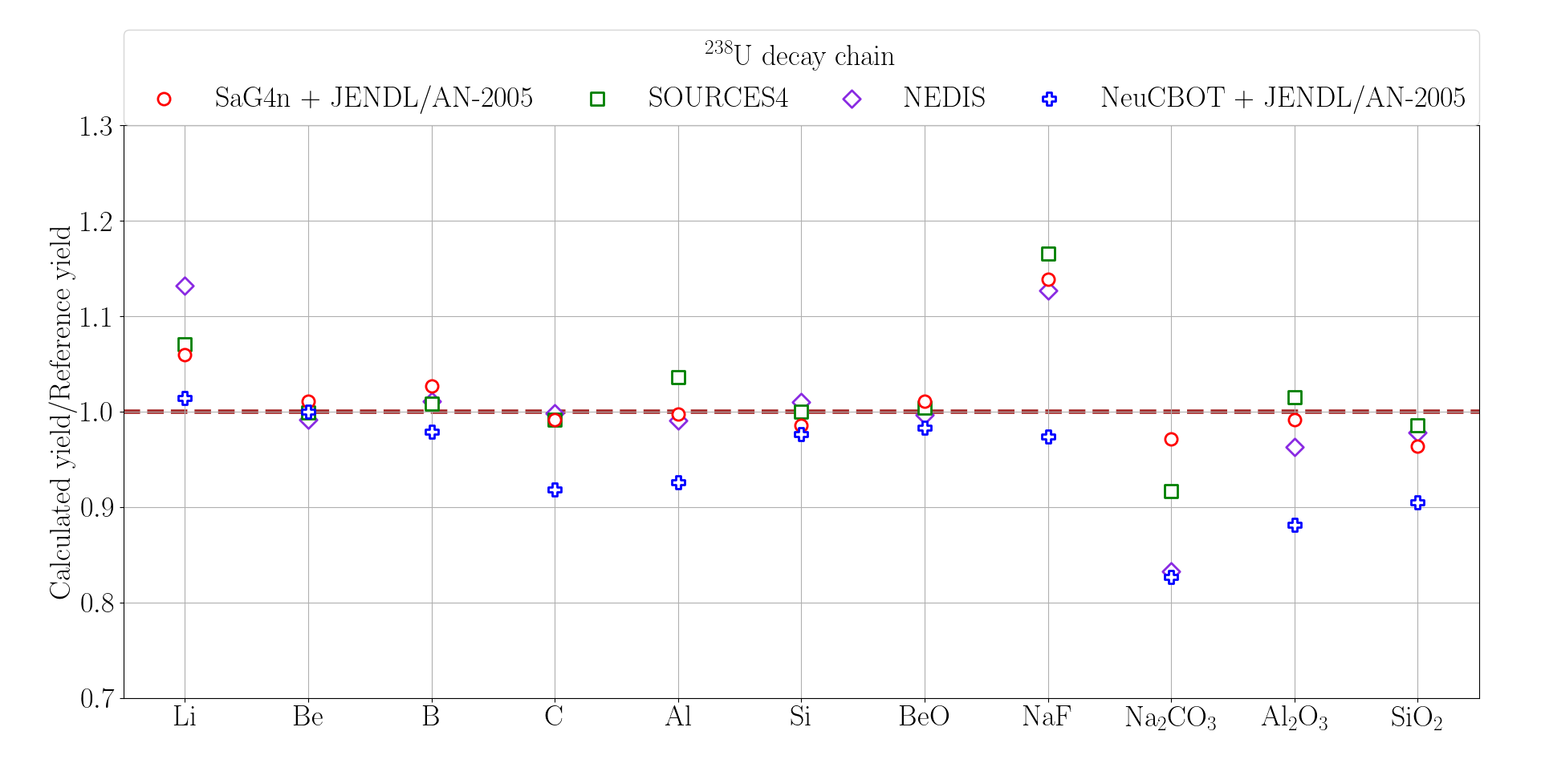}
    \caption{Comparison of neutron yields calculated for light nuclei by applying different codes, including the upgraded version of \NeuCBOT. The numerical values are normalized to evaluated data from various $\upalpha$-beam measurements \cite{fernandes} in order to demonstrate agreement between the calculations and measurements. The comparison in question is performed only for the uranium series.}
    \label{fig:comp_all_codes_238u}
\end{figure}

A detailed comparison between the codes is beyond the scope of this paper. Thick target neutron yields from beams of $\upalpha$-particles and naturally occurring radioactive decay chains from several different codes in comparison with experimental data have been published in Refs.~\cite{kudryavtsev_neutron_2020,Mendoza:2019vgf,westerdale_radiogenic_2017,NEDIS,fernandes,Gromov_2023}. Most codes are continuously developing and updated with new cross-sections so previous publications may not reflect the current status of the codes. We show here (Table~\ref{table:code-comparison}) a comparison between the 3 codes described above in calculating neutron yields from \ce{^238U}/\ce{^232Th} decay chains for widely used elements in detector components of low-background experiments. Fig. \ref{fig:comp_all_codes_232th} and Fig. \ref{fig:comp_all_codes_238u} present a comparison of neutron yield predictions for light nuclei derived from various computational tools, normalized to experimental data for the thorium and uranium decay chains, respectively; data from NEDIS \cite{nedisref}, another code that calculates neutron yields from \alphan\ reactions, are also included. While the agreement between different codes and experimental results is generally good for most of the materials analyzed, notable discrepancies are observed in specific cases, highlighting the need for further refinement of computational models and cross-section data.


%% file: TexFiles/Uncertanties.tex
\section{Typical uncertainties in neutron yield calculations}
\label{sec:Uncertainty}

Predicting the neutron fluxes induced by \alphan\ reactions in a material is a complex task, with uncertainties arising from multiple factors. A standardized methodology for evaluating \alphan\ reaction-related uncertainties is crucial to ensuring consistency across different computational approaches. Establishing common reference models and benchmark cases would significantly enhance the reliability of neutron yield predictions in various applications.

In low-background experiments, for instance, the neutron yield depends on factors such as the concentration of radiogenic contaminants in the detector components, the chemical (isotopic) composition of the materials, and the uncertainties associated with the calculation of the $\upalpha$-particle stopping power, as well as potential inaccuracies in cross-section data. For common nuclides and materials relevant to rare event search experiments, the uncertainty in the \alphan\ yield is typically within the range of \SI{30}{\percent}. However, in some cases, it can reach significantly higher values, up to $\mathcal{O}$(\SI{100}{\percent}) (e.g., $^{10}$B, $^{35}$Cl, $^{41}$K, $^{45}$Sc, $^{45}$Ti, $^{48}$Ti, $^{51}$V, $^{55}$Mn, $^{54}$Fe).

In the following, we discuss some of the main sources of uncertainty affecting \alphan\ yield calculations.

\subsection{Cross-sections: experimental data and nuclear model parameters}
\label{sec:Uncertainty:xsec}

The low efficiency in detecting neutrons and the challenges associated with conducting neutron spectroscopy experiments justify the limited availability of energy-dependent experimental \alphan\ cross-section data in the \EXFOR\ database for specific target nuclides, often lacking coverage across the entire energy range.

It is also possible that multiple measurements of the same nuclide often disagree with each other, 
possibly due to differences in the experimental setups or the corrections applied when interpreting
the results of the experiments. Furthermore, measurements catalogued by \EXFOR\ often have inconsistent treatments of experimental uncertainties, often accounting for varying levels of precision.

For what concerns the evaluated cross-section, the results of the nuclear reaction codes depend on the used models, the applied values for the model parameters, and even on the technical implementation of the models in code (e.g.\ on the binning of the level density tables \cite{Pereira2016}). Concerning the OMPs, the Reference Input Parameter Library for Calculation of Nuclear Reactions and Nuclear Data Evaluations (\RIPL)\footnote{Available at: \url{https://www-nds.iaea.org/RIPL-3}.} \cite{Capote2009}  generally recommends using individual-nucleus optical potentials where possible, while for nuclei far from the valley of stability, the use of microscopic optical potentials is recommended.

For \alphan\-reactions, the chosen $\upalpha$-OMP has the greatest impact on the obtained results \cite{Pereira2016}. \RIPL\ list in \cite{Capote2009} several global $\upalpha$-OMP suitable for $\upalpha$-induced reactions, see Table~\ref{table:models}. Out of these, the model \texttt{Avrigeanu2014} \cite{Avrigeanu:2014hma} is the most recent\footnote{Note that the different model \texttt{Avrigeanu1994} \cite{Avrigeanu1994} is only suitable for $\upalpha$ \emph{emission}, see \cite{Capote2009}.}. In addition, an $\upalpha$-OMP can be constructed from nucleon OMPs by Watanabe's folding approach \cite{Watanabe1958,Madland1988}.

\begin{sidewaystable}[!htbp]
    \centering
    \caption{\label{table:models} Nuclear models relevant for calculating the \alphan\ excitation function; a ``p'' indicates a phenomenological model, a ``m'' indicates a microscopic model. A hollow bullet (\textopenbullet) in the second or third column indicates if they are available in \TALYS\ or \EMPIRE, a filled bullet (\textbullet) indicate the default option, a square ($\square$) indicate the model can be loaded from the \RIPL\ database.}
    \begin{tabular}[t]{l @{\hskip 10pt} c @{\hskip 20pt} c @{\hskip 10pt} c l}
        \toprule
        \multirow{3}{*}{Model or data set} & \multirow{3}{*}{Type} & \multicolumn{2}{c}{Available in:} & \multirow{3}{*}{\hspace{20pt} Comments} \\ \cmidrule{3-4}
        & & TALYS & EMPIRE & \\
        & & 1.96  & 3.2    & \\
        \midrule
        Global $\upalpha$-OMP & & & & \\
        \hspace{24pt} \texttt{Avrigeanu2014} & p & \textbullet & & \hspace{20pt} \cite{Avrigeanu:2014hma,Avrigeanu2009}, $45 \leq A \leq 209$, $E \lesssim \SI{12}{\MeV}$ \\
        \hspace{24pt} \texttt{Kumar2006} & p & & \textbullet & \hspace{20pt} \cite{Kumar2006}, $12 \leq A \leq 209$, $E < \SI{140}{\MeV}$ \\
        \hspace{24pt} \texttt{Watanabe Folding Method} & p & \textopenbullet & & \makecell[lt]{\hspace{20pt} Folding \cite{Watanabe1958,Madland1988} of \texttt{KoningDelaroche2003} \cite{KoningDelaroche2003,Koning2021} \\ \hspace{20pt} nucleon potential, $24 \leq A \leq 209$, $\SI{2}{\keV} \leq E \leq \SI{200}{\MeV}$} \\
        \hspace{24pt} \texttt{Demetriou2002-1} & m & \textopenbullet & & 
        \hspace{20pt} \cite[table 1]{Demetriou2002}, volume term, $E \lesssim \SI{12}{\MeV}$\\
        \hspace{24pt} \texttt{Demetriou2002-2} & m & \textopenbullet & & \hspace{20pt} \cite[table 2]{Demetriou2002}, volume and surface term, $E \lesssim \SI{12}{\MeV}$\\
        \hspace{24pt} \texttt{Demetriou2002-3} & m & \textopenbullet & & \hspace{20pt} \cite{Demetriou2002}, dispersive potential, $E \lesssim \SI{12}{\MeV}$\\
        \hspace{24pt} \texttt{Strohmaier1982} & p & $\square$ & $\square$ & 
        \hspace{20pt} \cite{Strohmaier1982}, $40 \leq A \leq 100$, $\SI{1}{\MeV} \leq E \leq \SI{30}{\MeV}$\\
        \hspace{24pt} \texttt{McFadden1966} & p & \textopenbullet & $\square$ & 
        \hspace{20pt} \cite{McFadden1966}, $16 \leq A \leq 208$, $\SI{1}{\MeV} \leq E \leq \SI{25}{\MeV}$ \\
        \hspace{24pt} \texttt{Huizenga1962} & p & $\square$ & $\square$ & 
        \hspace{20pt} \cite{Huizenga1962}, $20 \leq A \leq 235$, $\SI{1}{\MeV} \leq E \leq \SI{46}{\MeV}$ \\
        Global nucleon-OMP & & & & \\
        \hspace{24pt} \texttt{Morillon2004} & p & $\square$ & $\square$ & 
        \hspace{20pt} \cite{Morillon2004,Morillon2006}, dispersive model, $27 \leq A \leq 209$, $\SI{1}{\keV} \leq E \leq \SI{200}{\MeV}$\\
        \hspace{24pt} \texttt{KoningDelaroche2003} & p & \textbullet & $\square$ & 
        \hspace{20pt} \cite{KoningDelaroche2003}, $24 \leq A \leq 209$, $\SI{1}{\keV} \leq E \leq \SI{200}{\MeV}$\\
        \hspace{24pt} \texttt{KoningDelaroche2003-dis} & p & \textopenbullet & & 
        \hspace{20pt} \cite{KoningDelaroche2003,Koning2021}, dispersive model, $24 \leq A \leq 209$, $\SI{1}{\keV} \leq E \leq \SI{200}{\MeV}$\\
        \hspace{24pt} \texttt{JLM-MOM} & m & \textopenbullet & & 
        \makecell[lt]{\hspace{20pt} \cite{Bauge2001} Jeukenne-Lejeune-Mahaux calculation with MOM code, \\ \hspace{20pt} $40 \leq A \leq 209$, $E \leq \SI{200}{\MeV}$}\\
        Nuclear mass & & & & \\
        \hspace{24pt} \texttt{AME2020} & p & \textbullet & & 
        \hspace{20pt} \cite{Huang2017}, the Atomic Mass Evaluation \\
        \hspace{24pt} \texttt{Goriely2016} & m & \textopenbullet & & 
        \hspace{20pt} \cite{Goriely2016}, Gogny-Hartree-Fock-Bogoliubov nuclear mass model\\
        \hspace{24pt} \texttt{FRDM2016} & m & \textopenbullet & & 
        \hspace{20pt} \cite{Moeller2016}, Finite-Range Droplet Macroscopic model\\
        \hspace{24pt} \texttt{Goriely2009} & m & \textopenbullet & & 
        \hspace{20pt} \cite{Goriely2009}, Skyrme-Hartree-Fock-Bogoliubov nuclear mass model\\
        \hspace{24pt} \texttt{FRDM1995} & p &  & \textbullet &  \hspace{20pt} \cite{MOLLER1995185},  Finite-Range Droplet Macroscopic model\\
        \hspace{24pt} \texttt{DZ} & m & \textopenbullet & & 
        \hspace{20pt} Duflo-Zuker mass formula, unpublished\\
        \makecell[lt]{Nuclear structure \\ (experimental data from RIPL-3 \cite{Capote2009} \\ and \ldots)} & & & &\\
        \hspace{24pt} \texttt{FRDM1995} & p &  & \textbullet &  \hspace{20pt} \cite{MOLLER1995185},  Finite-Range Droplet Macroscopic model\\
        \hspace{24pt} \texttt{Goriely2016} & m & \textopenbullet & & \hspace{20pt} \cite{Goriely2016}, Gogny-Hartree-Fock-Bogoliubov nuclear mass model\\
        \hspace{24pt} \texttt{Goriely2009} & m & \textbullet & & \hspace{20pt} \cite{Goriely2009}, Skyrme-Hartree-Fock-Bogoliubov nuclear mass model\\
        \bottomrule
    \end{tabular}
\end{sidewaystable}

\setcounter{table}{4}
\begin{sidewaystable}[!htbp]
    \centering
    \caption{\label{table:models2} Nuclear models relevant for calculating the \alphan\ excitation function; a ``p'' indicates a phenomenological model, a ``m'' indicates a microscopic model. A hollow bullet (\textopenbullet) in the second or third column indicates if they are available in \TALYS\ or \EMPIRE, a filled bullet (\textbullet) indicate the default option, a square ($\square$) indicate the model can be loaded from the \RIPL\ database. (Continued)}
    \begin{tabular}{l @{\hskip 10pt} c @{\hskip 20pt} c @{\hskip 10pt} c l}
        \toprule
        \multirow{3}{*}{Model or data set} & \multirow{3}{*}{Type} & \multicolumn{2}{c}{Available in:} & \multirow{3}{*}{\hspace{20pt} Comments} \\ \cmidrule{3-4}
        & & TALYS & EMPIRE & \\
        & & 1.96  & 3.2    & \\
        \midrule
        Level density & & & & \\
        \hspace{24pt} \texttt{EGSM} & p & & \textbullet & 
        \hspace{20pt} \cite{DArrigo1994}, Enhanced Generalized Superfluid Model\\
        \hspace{24pt} \texttt{GSM} & p & \textopenbullet & & 
        \makecell[lt]{\hspace{20pt} \cite{Ignatyuk1993}, Generalized Superfluid Model; \\\hspace{20pt} may include explicitly collective enhancement} \\
        \hspace{24pt} \texttt{BFM} & p & \textopenbullet & & \makecell[lt]{\hspace{20pt} \cite{Dilg1973,Grossjean1985,Demetriou2001}, Back-shifted Fermi gas Model; \\\hspace{20pt} may include explicitly collective enhancement}\\
        \hspace{24pt} \texttt{CTM} & p & \textbullet & \textopenbullet & 
        \makecell[lt]{\hspace{20pt} \cite{Gilbert1965}, Constant Temperature Model plus Fermi gas Model, \\ \hspace{20pt} TALYS and EMPIRE differ in the used parameterization, \\ 
        \hspace{20pt} TALYS uses the parameters of \cite{KONING200813}; \\ 
        \hspace{20pt} may include explicitly collective enhancement}\\
        \hspace{24pt} \texttt{Hilaire2012} & m & \textopenbullet & & 
        \hspace{20pt} \cite{Hilaire2012}, temperature-dependent Gogny-Hartree-Fock-Bogoliubov model\\
        \hspace{24pt} \texttt{Goriely2008} & m & \textopenbullet & \textopenbullet & \hspace{20pt} \cite{Goriely2008}, deformed Skyrme-Hartree-Fock-Bogoliubov model\\
        \hspace{24pt} \texttt{Demetriou2001} & m & \textopenbullet & & \makecell[lt]{\hspace{20pt} \cite{Demetriou2001}, Hartree-Fock-BCS model; misidentified in \TALYS' manual as \\ \hspace{20pt} \texttt{Goriely2001} \cite{Goriely2001}}\\
        \makecell[lt]{Photon strength function \\ (for E1 unless otherwise stated)} & & & &\\
        \hspace{24pt} \texttt{SMLO} & p & \textbullet & & \makecell[lt]{\hspace{20pt} \cite{Goriely2019}, Simplified Modified Lorentzian model for E1 and \\ \hspace{20pt} standard Lorentzian for M1 with low-energy upbend} \\
        \hspace{24pt} \texttt{MLO3} & p & & \textopenbullet & 
        \hspace{20pt} \cite{Plujko2002}, Modified Lorentzian model \\
        \hspace{24pt} \texttt{MLO2} & p & & \textopenbullet & 
        \hspace{20pt} \cite{Plujko2002}, Modified Lorentzian model \\
        \hspace{24pt} \texttt{MLO1} & p & & \textbullet & 
        \hspace{20pt} \cite{Plujko2002}, Modified Lorentzian model\\
        \hspace{24pt} \texttt{GFL} & p & & \textopenbullet & 
        \hspace{20pt} \cite{Mughabghab2000}, Generalized Fermi Liquid model \\
        \hspace{24pt} \texttt{EGLO} & p & & \textopenbullet & 
        \hspace{20pt} \cite{Kopecky1993}, Enhanced Generalized Lorentzian model\\
        \hspace{24pt} \texttt{GLO} & p & \textopenbullet & \textopenbullet & \hspace{20pt} \cite{Kopecky1990}, Generalized Lorentzian model\\
        \hspace{24pt} \texttt{SLO} & p & \textopenbullet & \textopenbullet & \makecell[lt]{\hspace{20pt} \cite{Brink1957,Axel1962}, Standard Lorentzian model; used for M1 (default for \\ \hspace{20pt} EMPIRE) and $Xl,\;l>1$ (default for both codes), \\ \hspace{20pt} with parameters from RIPL-3 \cite{Capote2009}}\\
        \bottomrule
    \end{tabular}
\end{sidewaystable}

\setcounter{table}{4}
\begin{sidewaystable}[!htbp]
    \centering
    \caption{\label{table:models3} Nuclear models relevant for calculating the \alphan\ excitation function; a ``p'' indicates a phenomenological model, a ``m'' indicates a microscopic model. A hollow bullet (\textopenbullet) in the second or third column indicates if they are available in \TALYS\ or \EMPIRE, a filled bullet (\textbullet) indicate the default option, a square ($\square$) indicate the model can be loaded from the \RIPL\ database. (Continued)}
    \begin{tabular}{l @{\hskip 10pt} c @{\hskip 20pt} c @{\hskip 10pt} c l}
        \toprule
        \multirow{3}{*}{Model or data set} & \multirow{3}{*}{Type} & \multicolumn{2}{c}{Available in:} & \multirow{3}{*}{\hspace{20pt} Comments} \\ \cmidrule{3-4}
        & & TALYS & EMPIRE & \\
        & & 1.96  & 3.2    & \\
        \midrule
        \hspace{24pt} \texttt{HFB-QRPA-D1M} & m & \textopenbullet & & 
        \makecell[lt]{\hspace{20pt} \cite{Goriely2018}, D1M Gogny-Hartree-Fock-Bogoliubov model plus\\
        \hspace{20pt} Quasiparticle Random Phase Approximation (QRPA), \\ 
        \hspace{20pt} E1 and M1 parameterization with low-energy upbend}\\
        \hspace{24pt} \texttt{Daoutidis2012} & m & \textopenbullet & & 
        \hspace{20pt} \cite{Daoutidis2012}, temperature-dependent relativistic mean field model\\
        \hspace{24pt} \texttt{HFB-QRPA} & m & \textopenbullet & & 
        \hspace{20pt} \cite{Goriely2004}, Skyrme-Hartree-Fock-Bogoliubov model plus QRPA\\
        \hspace{24pt} \texttt{HFBCS-QRPA} & m & \textopenbullet & & 
        \hspace{20pt} \cite{Goriely2002}, Skyrme-Hartree-Fock BCS model plus QRPA\\
        \hspace{24pt} \texttt{Goriely1998} & & \textopenbullet & & 
        \hspace{20pt} \cite{Goriely1998}, hybrid model\\
        Width fluctuation correction & & & & \\
        \hspace{24pt} \texttt{GOE} & & \textopenbullet & & 
        \hspace{20pt} \cite{Verbaarschot1985}, Gaussian Orthogonal Ensemble of Hamiltonian matrices-model\\
        \hspace{24pt} \texttt{HRTW} & & \textopenbullet & \textbullet & 
        \hspace{20pt} \cite{Hofmann1980}, Hofmann-Richert-Tepel-Weidenm\"uller model\\
        \hspace{24pt} \texttt{Moldauer1980} & & \textbullet & & 
        \makecell[lt]{\hspace{20pt} \cite{Moldauer1980}, Approximate method for practical applications that does not\\ \hspace{20pt} include direct-compound reaction interference}\\
        \bottomrule
    \end{tabular}
\end{sidewaystable}

From the newer global, phenomenological nucleon-nucleus OMP contained in \RIPLv{3} \cite{Capote2009}, only two, \texttt{KoningDelaroche2003} \cite{KoningDelaroche2003} and \texttt{Morillon2006} \cite{Morillon2004,Morillon2006}, covers partially our nuclide and energy ranges of interest. The latter is a dispersive model and also of \texttt{KoningDelaroche2003} exists an unpublished dispersive version, which is provided by \TALYSv{1.96} \cite{Koning2021}. \TALYS\ provides also local OMPs covering some of our nuclides of interest \cite{KoningDelaroche2003}; further local OMP are listed in \cite{Capote2009} and \cite[Annex 5.D]{IAEA2006}. It also offers the option to use a microscopic nucleon OMP, the JLM model of Bauge \emph{et al.} \cite{Bauge2001}, that partially covers the relevant nuclides.
Both \TALYS\ and \EMPIRE\ can read OMPs from \RIPL; in addition \TALYS\ provides the $\upalpha$-OMPs \texttt{Avrigeanu2014} \cite{Avrigeanu:2014hma,Avrigeanu2009} and \texttt{Demetriou2002-1} \cite[table 1]{Demetriou2002} which are not part of \RIPLv{3} and finally the possibility to read in tabulated OMPs from a user-provided file.

Further properties that affect the results are nuclear mass, nuclear structure, level density, and photon strength function; whereas the impact of the width fluctuation correction is only minor \cite{Pereira2016}. An overview of available models is given e.g.\ in \RIPL~\cite{Capote2009}. For several combinations of target nuclides and projectiles, the \TENDL\ library provides also ``best'' settings and error estimations for these properties, which can be used in \TALYS~\cite{koning_tendl_2019,Koning2021}. In the current \TALYS\ version, ``best'' settings are mostly available for incident n and $\upgamma$ and not for $\upalpha$; in this case, \TALYS\ falls back to its default settings and their uncertainties.

Table~\ref{table:models} summarize the relevant models and data sets for the calculation of the \alphan-excitation function. Uncertainties on the calculation are caused by the propagation of the uncertainties on the model parameter whereas the selection of a particular, inaccurate model may cause a systematic uncertainty. One has also to note that for the ten lightest nuclides in Table~\ref{tab:xsdata} calculations with \TALYS\ or \EMPIRE\ are not suitable because the nuclide masses are below the applicability range of any of the listed OMPs. At this mass range, R-matrix theory may be a more suitable approach than the statistical treatment. An overview of suitable R-matrix codes is given in \cite{Leeb2017,Thompson2019}.

\begin{figure}[!htbp]
\centering
\includegraphics[width=0.9\textwidth]{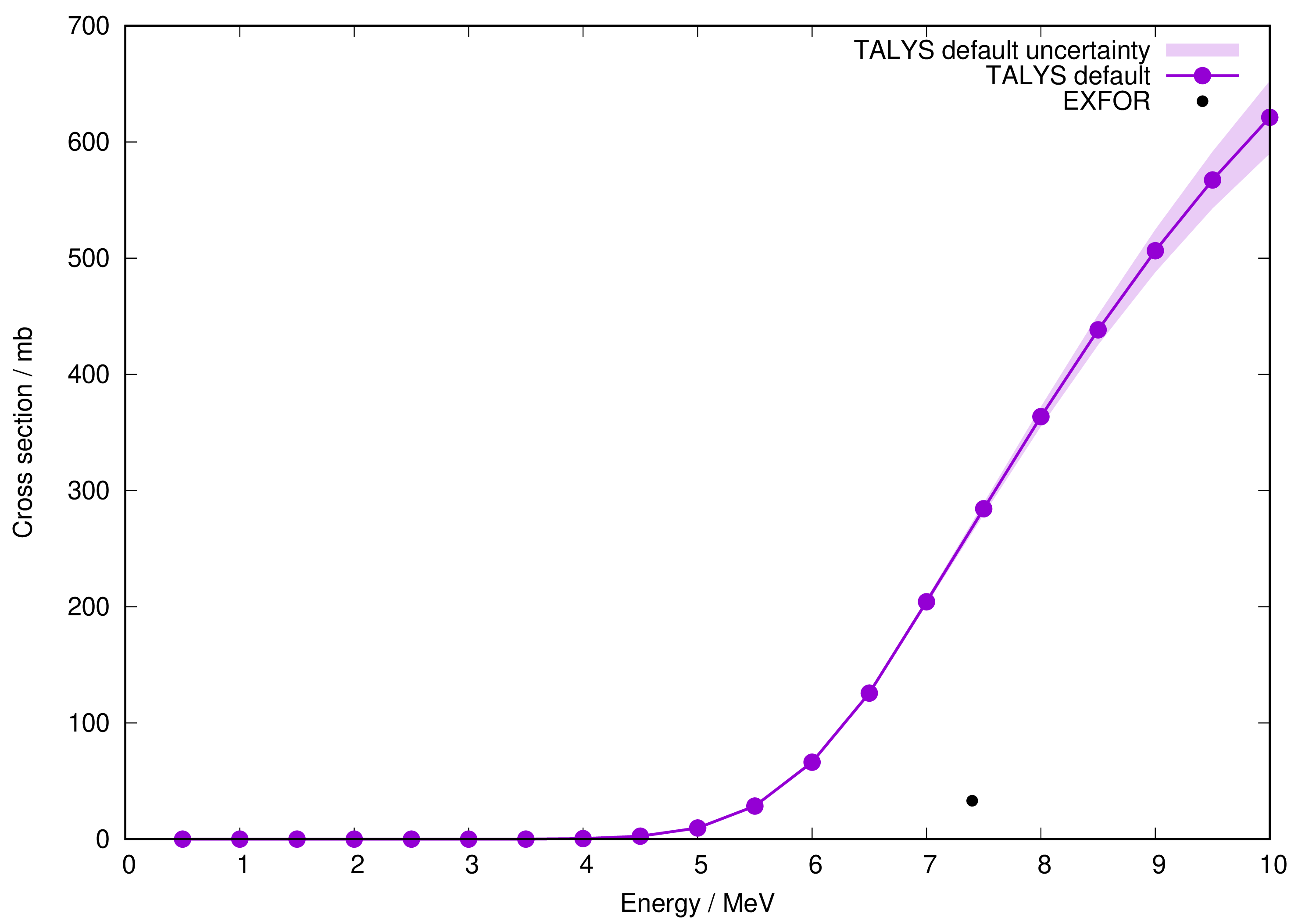}
\caption{\label{fig:talysAr40}Excitation function for $\mathrm{^{40}Ar(\upalpha, n)}$ (\emph{dark violet line}) as calculated with \TALYSv{1.96} based on default settings, see also table~\ref{table:models}. The associated uncertainty (\emph{light violet band}) is based on sampling the input parameter space with \TASMAN: 2750 samples within the parameter's uncertainties were drawn. For comparison, the only existing measurement by Schwartz \emph{et al.} \cite{Schwartz1956} from the \EXFOR\ data base is shown (\emph{black data point}).}
\end{figure}

With the auxiliary tool \TASMAN\footnote{Currently available in version 2.0 at: \url{https://nds.iaea.org/talys}}, it is possible to perform random sampling of the input parameters space within their respective uncertainties and hence propagate these uncertainties to the result of \TALYS~\cite{koning_tendl_2019}\footnote{Within the TALYS-related literature, this approach is called \emph{Total Monte Carlo} \cite{koning_tendl_2019}.}, i.e.\ in our case the excitation function of \alphan\ reactions. In Fig.~\ref{fig:talysAr40}, as an example, the excitation function for $\mathrm{^{40}Ar(\upalpha, n)}$, calculated with \TALYSv{1.96} using default settings, is shown along with its associated uncertainty. For comparison, the only existing measurement by Schwartz \emph{et al.} \cite{Schwartz1956} is also displayed. It is worth noting that this data point was derived from a single scattering angle and extrapolated to the total cross-section, introducing significant uncertainties, which likely account for its deviation from the theoretical calculations.

\subsection{Assay results}
\label{sec:Uncertainty:assay}

The source of the neutrons generated in the \alphan\ reactions are $\upalpha$-emitters coming mostly from the \ce{^238U}-chain. The activity concentrations must be
measured for all relevant materials and specific nuclides, or at least for sub-chains, which are in equilibrium - see the discussion in Section~\ref{sec:Inputs:Isotopes:LowBG} and Fig.~\ref{fig:U_chain}. The activities in the upper part of the \ce{^238U}-chain, assayed with the help of mass spectrometry (ICP-MS), are determined with a precision of about \SIrange{10}{40}{\percent}. It is assumed here that the measurements are performed with the highest sensitivities (sub 0.1 ppt $\simeq$ \SI{1}{\uBq\per\kg}). ICP-MS may be also used to investigate the high-purity material contamination not only with radioactive nuclides, but also with various elements down to \num{e-9} g/g~\cite{Ferella2021}. The typical uncertainty of such measurements is about \SI{30}{\percent}. This may be interesting especially if elements with high cross-section for \alphan\ reactions are considered. ICP-MS may also be applied to determine the isotopic composition (stable and long-lived isotopes) of the material in question. Depending on the isotopes, the precision of its abundance determination may vary from \SIrange{1}{30}{\percent}~\cite{Ferella2021}.

Measurements of \ce{^226Ra}, \ce{^222Rn} and its short-lived daughters (down to \ce{^214Po}, middle part of the \ce{^238U}-chain, see Fig.~\ref{fig:U_chain}) close to the detection limits (\SI{\approx 10}{\uBq\per\kg}) are also performed with a typical precision of about \SI{30}{\percent}~\cite{Neder:2000}.

The lower part of the \ce{^238U}-chain (\ce{^210Pb}--\ce{^210Po}) is assayed with the lowest sensitivity of about \SI{1}{\milli\becquerel\per\kg}. The procedure requires extensive chemistry to separate \ce{^210Po} from the sample matrix, deposit it on a dedicated disc, and count the activity. A time series of \ce{^210Po} measurements allows to determine the \ce{^210Pb} activity concentration with the precision of about \SIrange{15}{20}{\percent}~\cite{Mroz2021}. 

To conclude, one can state that the measurements of the activity concentrations of $\upalpha$-emitters are performed with a precision of
about \SI{30}{\percent}. One should underline here that this concerns tests performed with the highest available sensitivities and applying various techniques (mass spectrometry, gamma counting, chemical separation of \ce{^210Po}). If lower sensitivities are required the precision is higher, in average \SIrange{15}{20}{\percent} depending on the sample and technique.

\subsection{Material composition}
\label{sec:Uncertainty:comp}

In the assessment of potential neutron backgrounds for next-generation rare event search detectors, the uncertainty in the chemical composition of materials is frequently overlooked. Typically, detector material assays focus primarily on measuring the radioactivity levels of uranium and thorium chains. Nevertheless, as introduced in Section \ref{sec:Inputs:Isotopes:LowBG}, the material chemical composition should always be investigated in detail and assayed when necessary to accurately calculate the \alphan\ neutron yield. The uncertainty in neutron production due to variations in elemental composition can be significant, and in some cases, comparable to or even larger than the uncertainty in the assay results or the \alphan\ cross-section. 

In the best-case scenario, the chemical composition of a component is provided by the manufacturer, often with limited or no information on the associated uncertainties. In the worst-case scenario, when the composition is unknown, an elemental analysis of the material must be conducted, typically achieving a precision of about \SIrange{15}{20}{\percent}.

Electronic circuit components, such as common resistors, provide an excellent example of materials with non-homogeneous and challenging compositions. Variations in the placement of layers of borosilicate glass within the small simulated volume can significantly impact the neutron yield. 

Elements such as aluminum, fluorine, and boron, commonly found in electronic components, contribute significantly to neutron production due to their large \alphan\ cross-sections. Even when present in small fractions, these elements can account for several percent of the total neutron yield. Consequently, variations in their mass fraction within the material contribute substantially to the overall uncertainty in neutron yield calculations.

As an example, the study of resistors used in the light readout system of a low-background experiment is presented in Table \ref{table:resistor_composition}. The chemical elements are listed based on their contribution to neutron yield, and the nominal mass fraction provides the most accurate input for simulations. As shown in Table \ref{table:resistor_composition}, although aluminum constitutes 33\% of the mass fraction, it contributes approximately 68\% of the total neutron yield, highlighting its substantial impact. Similarly, boron, which represents only 1\% of the mass fraction, accounts for about 17\% of the neutron yield due to its high \alphan\ cross-section. If the mass fraction of boron were increased to 2\% or 3\%, the total neutron yield would rise by approximately 17\% or 34\%, emphasizing its critical role in the neutron yield balance.

These findings demonstrate that propagating uncertainties in the mass fractions of elements into neutron yield calculations is not straightforward and requires careful consideration. Another critical aspect is the uniformity of the chemical composition within the material. For significant contributors to neutron yield, extensive elemental analyses may be required to determine the distribution of chemical elements within the component.

\begin{table}[ht]
\caption{The chemical composition of a resistor is given with the element's mass fraction as assumed in the MC simulation model and calculated expected neutron yield due to \alphan\ reactions. The neutron production is calculated based on the \SaGFN\ simulation and measured radioactivity level of uranium and thorium chains.}
\label{table:resistor_composition}
\centering
\begin{tblr}{
	colspec={
	c
	Q[si={table-format=1.2,table-number-alignment=center},c]
	Q[si={table-format=1.2e-2,table-number-alignment=center},c]
	},
	row{1}={guard}
}
\toprule
Element & Nominal mass fraction & Neutron yield / \si{\per\kg\per\second}\\\midrule
Al & 0.33 & 3.21e-05 \\
B & 0.01 & 8.30e-06 \\
O & 0.55 & 3.07e-06 \\
Mg & 0.01 & 7.06e-07 \\
Si & 0.04 & 3.82e-07 \\
Ca & 0.01 &  7.53e-09 \\
Ni & 0.02 & 1.11e-09 \\
Cu & 0.01 & 5.52e-10 \\
\bottomrule
\end{tblr}
\end{table}


%% file: TexFiles/Measurements.tex
\section{Data needs}
\label{sec:Measurement}

The success of the accurate \alphan\ neutron yield calculation for spent fuel management, low-background experiments, and nuclear astrophysics heavily relies on accurate experimental data and nuclear models. Unfortunately, much of the existing experimental data is outdated, incomplete, and characterized by significant or inconsistent uncertainties in the cross-sections. Furthermore, experimental data on neutron emission angular distributions are rare, and information regarding partial cross-sections and correlated $\gamma$-ray emission is even more scarce. Consequently, the evaluated nuclear libraries suffer from incompleteness (i.e. only a few nuclides are available) or reliance on outdated evaluations. Calculations involving numerous nuclides require combining evaluated cross-section files, relying on experimental data, with theory-driven cross-section files or nuclear models.

Recognizing the urgency to update nuclear data libraries for charged-particle-induced reactions, and \alphan\ in particular, the IAEA's Nuclear Data Section (NDS) has initiated a global collaborative effort \cite{IAEA2021, IAEA2023}. This initiative aims to produce updated and reliable data for charged-particle-induced reactions. 
Notably, for compounds such as UF$_{6}$ and PuF$_{4}$ which are involved in processes such as uranium enrichment, storage of depleted uranium, and pyrochemical reprocessing, discrepancies between new evaluations and the 1991 reference data reached the 25-50 \% level, emphasizing the critical need for accurate $^{19}$F\alphan\ reaction cross-sections.

New massive argon-based detectors for rare event searches have been proposed or are in the construction phase \cite{DarkSide-20k:2017zyg,GlobalArgonDarkMatter:2022ppc}. For these experiments, the direct measurement of the \alphan\ cross-section on $^{40}$Ar is crucial for accurate background calculations, as experimental data for this element are either very old or essentially absent in the large part of the energy range of interest.

The imperative for enhanced experimental and evaluated data libraries extends beyond the cases of  $^{19}$F and $^{40}$Ar, encompassing \alphan\ reactions across various light and medium-light elements essential for diverse applications. The data available in the \EXFOR\ database show large inconsistencies with respect to the reported uncertainties, and \alphanx\ and \alphang\ cross-sections data are particularly limited. Furthermore, the development of nuclear reaction codes, particularly those employing R-matrix and statistical models, as well as source codes used for calculating neutron sources from existing cross-section and stopping-power data, necessitates continuous updates.


To adequately address these demands for new measurements, improved data libraries, and sophisticated software, an international collaboration is paramount. Coordinated efforts among interdisciplinary groups actively involved in \alphan\ studies can substantially enhance our comprehension of this reaction and match the nuclear data needs in different fields of science and applications.

%% file: TexFiles/Conclusions.tex
\section{Conclusions}
\label{sec:Concl}

Accurate predictions of \alphan\ neutron yields are critical for reducing uncertainties in sensitivity calculations for next-generation physics experiments operating in the keV–-MeV energy range. A detailed understanding of the mechanisms underlying radiogenic neutron production is equally crucial for advancing nuclear astrophysics, refining nuclear technologies —including enrichment processes and the development of advanced reactor systems such as molten salt reactors (MSRs)—and supporting applications in fields such as medical physics.

This paper provides a comprehensive review of the current state of \alphan\ yield calculations, presenting the most advanced computational models, experimental cross-section databases, and state-of-the-art computational tools.

The main features of the most commonly used computational tools for neutron yield calculations are presented, providing a comparative analysis across various databases.  While agreement between different codes and experimental results is generally good for most materials analyzed, deviations persist in specific cases. This underscores the importance of further developing and validating more accurate computational methods to reduce uncertainties and improve the reliability of predictions.

This review also discusses the key uncertainties in neutron yield predictions, including those arising from uranium and thorium contamination measurements, material chemical composition, cross-section data, and theoretical models. Overall, significant uncertainties in \alphan\ yield predictions have been highlighted, partially driven by missing cross-section measurements, limited energy coverage, or inconsistencies among different experimentally measured cross-sections. For many nuclides and materials, the uncertainties in neutron yield calculations are typically estimated to be within 30\%. In some cases, the discrepancies can be much larger due to the absence of experimental \alphan\ cross-section data or significant variability in the available datasets.

These findings highlight not only the need to refine models and improve theoretical cross-section calculations, but also the necessity of expanding experimental programs to include cross-section measurements for a broader range of materials and energy regimes. Addressing these gaps is crucial to ensuring more reliable cross-section data. Further efforts should focus on obtaining experimental \alphan\ cross-sections for mid-Z materials, which are currently underrepresented in databases. Priorities for new measurements have been identified, particularly for materials commonly used in rare event physics experiments, where accurate neutron yield predictions are essential.

Moreover, the importance of improving the understanding of correlated $\gamma$-ray emissions and \ngamma\ cross-sections has been emphasized. These decay data remain critical for interpreting detector triggers and vetoes in rare event search experiments, which are critical aspects of low-background detectors.

%% file: TexFiles/Acknowledgments.tex
\section*{Acknowledgments}

The CIEMAT group is funded by the Spanish Ministry of Science (MICINN) through the grants PID2022-138357NB-C22 and PID2021-123100NB-I00, and by the project MCINN/AEI/1013039/   501100011033/FEDER, UE. V.P. is additionally supported by the grant 2018-T2/TIC-10494 of the Community of Madrid. This work was also funded in the framework of the Polish NCN Opus research grant (UMO2019/33/B/ST2/02884).  V.K. is supported by STFC (grants ST/W000547/1, ST/Y004841/1, ST/S003398/1) and the University of Sheffield. S.W. is sponsored by the University of California Riverside (USA) and the National Science Foundation (NSF-PHYS 2310091).  H. K. is funded by the Austrian science fund (FWF) through project No. P 34778-N ELOISE. M.P. and I.L. acknowledge financial support from the contract no. 04/2022, Programme 5, Module 5.2 CERN-RO.
A.K. is supported by the Fermi Research Alliance, LLC under Contract No. DE-AC02-07CH11359 with the U.S. Department of Energy, Office of Science, Office of High Energy Physics. V.L. is supported by the Fundação para a Ciência e a Tecnologia FCT-Portugal (grant 2021.01039.CEECIND). M.G. acknowledges the  grant of the Russian Science Foundation (RSF) No. 24-12-00046.